\documentclass{intech_SIDRA}

\usepackage{color}
\usepackage{natbib}            
\usepackage{graphicx}          
\usepackage{amsmath} 
\usepackage{amssymb}
\usepackage{mathptmx}
\usepackage{times}
\usepackage{psfrag}
\usepackage[ansinew]{inputenc}

\chaptertitle{Analysis, Dimensioning and Robust Control of Shunt Active Filter for Harmonic Currents Compensation in Electrical Mains}

\authors{Andrea Tilli, Lorenzo Marconi, Christian Conficoni}
\affiliation{Center for Complex Automated Systems (CASY),\\ 
Dept. of  Electronics, Computer  Engineering and Systems (DEIS),\\
University of  Bologna, \\ Viale Risorgimento 2, 40136 Bologna}
\country{Italy}

\begin{document}
\maketitle

\emph{\large Abstract} 
 \vspace{4 pt}
 \\
In this chapter some results related to Shunt Active Filters (SAFs) and obtained by the authors and some coauthors are reported.
SAFs are complex power electronics equipments adopted to compensate for current harmonic pollution in electric mains, due to nonlinear loads.
By using a proper "floating" capacitor as energy reservoir, the SAF purpose is to inject in the line grid currents canceling the polluting harmonics.   
Control algorithms play a key role for such devices and, in general, in many power electronics applications.
Moreover, systems theory is crucial, since it is the mathematical tool that enables a deep understanding of the involved dynamics of such systems, allowing a correct dimensioning, beside an effective control. As a matter of facts, current injection objective can be straightforwardly formulated as an output tracking control problem. In this fashion, the structural and insidious marginally-stable internal/zero dynamics of SAFs can be immediately highlighted and characterized in terms of sizing and control issues.\\
For what concerns the control design strictly, time-scale separation among output and internal dynamics can be effectively exploited to split the control design in different stages that can be later aggregated, by using singular perturbation analysis. In addition, for robust asymptotic output tracking the Internal Model Principle is adopted.
\vspace{4 pt}
\\
In authors' opinion, SAF case is an illustrative example of common issues in dimensioning and control of complex power electronics equipments, hence the proposed design approach can be generalized for such class of systems (e.g. large-power electric drives; generators and converters, particularly for renewable energies; Uninterruptible Power Supplies, UPS, and power supplies for special applications as particle accelerators).  
Remarkable the role of "system theory approach" in enlightening crucial sizing issues, this fact strongly testifies how relevant the "control viewpoint" is in all the fields of engineering, particularly when complex dynamic behavior is requested.
\vspace{4 pt}
\\
The presented chapter has been invited for possible publication (after a review process) in the book "Robust Control / Book 1", ISBN 978-953-307-421-4, by INTECH (www.intechweb.org).\\

\section{Introduction}
Harmonic pollution in the AC mains determines additional power losses and may cause malfunctioning or even damage to connected equipments. Distortion of the currents circulating on electric mains is mainly originated by non linear loads, as AC/DC uncontrolled rectifiers used for motor drives, that absorb undesired current harmonics. Therefore, local countermeasures have to be taken in order to keep the portion of grid affected by distortion as small as possible, hence preventing relevant power losses and ``saving'' other equipments, connected to the rest of the grid.\\
Traditionally, passive filtering components have been adopted to cope with harmonic compensations, however they are affected by several drawbacks; they are very sensitive to network impedance variation and environmental conditions, moreover they need to be tuned on
fixed frequencies. In order to overcome those limitations, in the last decades, thanks also to the fast growth in power electronics and control processor technologies, a remarkable research attempt has been devoted to the study of the so-called Active Power Filters (APFs), both from a theoretical and technological point of view (see \cite{Gyugy76}, \cite{Akagy96}, \cite{Singh99}). These devices are able to properly work in a wide range of operating conditions, providing better performance and overtaking intrinsic limitations of passive devices, they are far more insensitive to network impedance, they can be tuned onto different frequencies just varying some software parameters. Furthermore, the system reliability is improved, resonance phenomena are avoided and a diagnosis system can be implemented on the control processor to monitor the system variables and adopt some recovery strategy in case of faulty conditions.

\begin{figure}[t]	
\centering
\includegraphics[width=8cm, angle=-90]{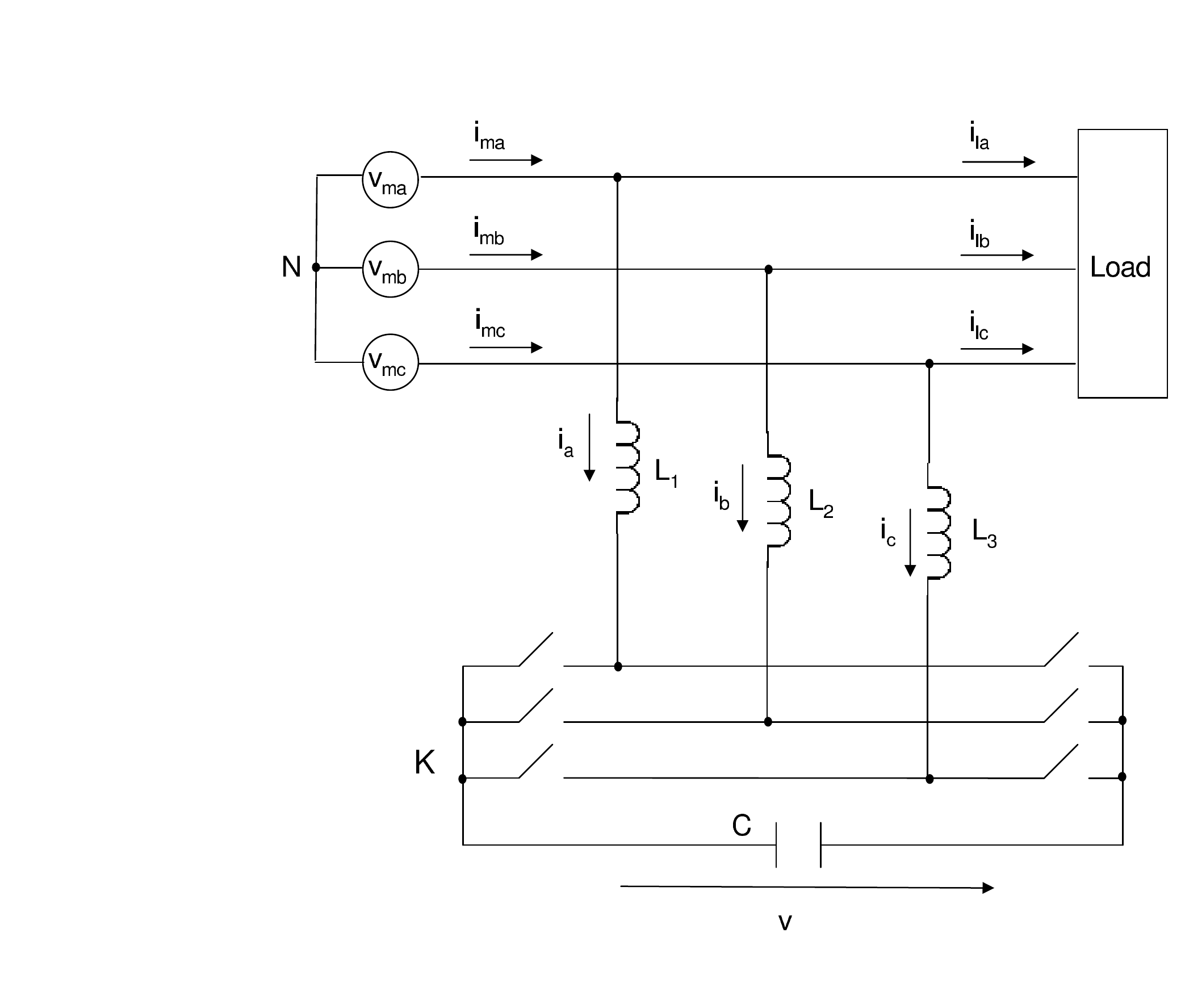} 
\caption{Shunt Active Filter scheme.}\label{fig:SAFscheme}
\end{figure}

In this chapter, the general issues related to analysis, dimensioning and control of a particular class of APFs, the so-called Shunt Active Filters (SAFs), are addressed; the main purpose of this kind of power system is to inject into mains a proper current, in order to cancel out, partially or totally, the power distortions generated by nonlinear loads. The SAFs considered in this work are based on a three-phases three-wires AC/DC boost converter topology (see Fig. \ref{fig:SAFscheme}) connected in parallel to the distorting loads.\\
The first step to properly design a SAF is the selection of suitable hardware components; as it will become clear in the next section, owing to the structural properties of the system, the sizing procedure cannot be considered apart from the canonical control aspects, hence a correct dimensioning algorithm \citep{Ronchi2002} is proposed to ensure feasibility of the desired control objectives and to minimize costs. In addition, according to such method, it is shown how a time-scale separation between different dynamics of SAF usually takes place ``for free''. This point is very useful for control design and stability analysis.\\
Once a correct hardware sizing has been carried out, the first control issues to deal with are: the current/power control algorithm and the load current analysis method adopted to define the filter current reference. Various solutions have been proposed in literature. As regards current/power harmonic tracking, in \citep{Chandra2000} an hysteresis current control \citep{Kazmierkowski1998} is proposed, while in \citep{Jeong1997} predictive current control is adopted. For what concerns the generation of the filter currents reference, beside Fast Fourier Transform Techniques, instantaneous power theory \citep{Akagi1984}, time domain correlation techniques \citep{VanHarmelen1993}, notch filter theory \citep{Rastogi1995} and other methods have been proposed. Solution based on state observer have been proposed, too, as in \citep{Bhattacharya1995} and \citep{Tilli2002}.\\
However, what renders the SAF control problem challenging and different from other conventional tracking problems is the presence of peculiar and unstable internal dynamics, given by the voltage dynamics of the DC-link capacitor bank. This capacitor bank is the main energy storage element, which provides the voltage, modulated by the control, to steer the filter currents and, at the same time, is required to oscillate to exchange energy with the line and the load to compensate for current harmonics.
Actually, this element needs to be carefully considered also in the previously-mentioned dimensioning stage; a correct capacitor sizing is crucial for control objective feasibility, whatever control technique is adopted.
Moreover, it can be shown that, if perfect harmonic compensation is achieved, the DC-link voltage dynamics are unstable, due to the system parasitic resistances that lead to a slow discharge of the capacitor.
Hence, a suitable stabilizing action for DC-link voltage dynamics needs to be provided. Since no additional circuit is used to feed the DC-link capacitor independently of the three-phase port used to inject currents, (see Fig. \ref{fig:SAFscheme}), the voltage stabilization would need to be integrated with the controller devoted to harmonic compensation (the AC/DC boost-based SAF is an underactuated system). This is a crucial point and it has to be tackled preserving harmonic compensation performances as far as possible.\\
In this work a power/current controller, based on Internal Model Principle, (see \cite{Marconi2003}, \cite{Marconi2004}, \cite{Marconi2007}) is designed in order to cancel current harmonics, ensuring robustness with respect to SAF parameter uncertainties. By exploiting the internal model approach, the proposed solution also allows to merge and solve at the same time the two above-mentioned problems of current harmonics isolation and current reference tracking.
As regards the robust stabilization of the DC-link voltage internal dynamics, a cascade control structure is proposed. An additional voltage controller, acting on the references of the power/current controller, is introduced. This controller is designed taking into account the structural voltage oscillations required for harmonic compensation and minimizing the impact on harmonic compensation. In particular, by exploiting a proper averaging \citep{Sanders1991} of the capacitor voltage dynamics, the average value of the capacitor voltage is chosen as output variable to be controlled \citep{Hanschke2006}.\\
As far as the overall stability is concerned, the previously mentioned time-scale separation between portions of SAF dynamics can be effectively exploited to decouple power/current tracking and voltage stabilization control problems, using \emph{averaging} and \emph{singular perturbation theory} techniques \citep{Khalil96}.

This chapter is organized as follows. In Section \ref{sec:modello}, the general framework is described, the SAF model is derived and the control objectives are formally defined. In Section \ref{sec:dimensionamento}, two methodological approaches are presented for the SAF components sizing. The first one is based on the knowledge of the load currents harmonic spectrum, the values selected for the hardware components are the minimums allowing the SAF to deal with the considered load distortion. Differently, the second approach is related to the maximum current of the AC/DC boost switching devices. In this case the selected components values are the minimums which enable the SAF to compensate for all possible loads giving distorted currents smaller or equal to the switches peak value. In Section \ref{sec:controllo} both the internal model-based power/current controller and the averaging voltage controller design are presented, stability analysis is carried out relying upon the time-scale separation imposed by the design algorithm; both the power and the voltage subsystem are proven to be asymptotically stable, then practical stability of the overall system is claimed exploiting general results on two time-scale averaged systems \citep{Teel2003}. The effectiveness of the proposed control solution is tested in Section \ref{sec:simulazioni} through simulations.
\section{Shunt active filter model and control problem statement} \label{sec:modello}
The scheme of the shunt active filter considered in this chapter is reported in Fig. \ref{fig:SAFscheme}, as mentioned in the introduction it is based on a three-phase three-wire AC/DC boost converter, where the main energy storage element is a DC-bus capacitor, while the inductances are exploited to steer the filter currents by means of the converter voltages. The switching devices of the three-leg bridge (also called "`inverter"') are usually realized by IGBTs (Insulated Gate Bipolar Transistors) and free-wheeling diodes.\\
In this work the following notation is used to denote the SAF variables; $v_{mabc}$=$(v_{ma}, v_{mb}, v_{mc})^T$ is the mains voltage sinusoidal balanced and equilibrated tern, $i_m$=$(i_{ma}, i_{mb}, i_{mc})^T$ are the mains currents, $i_l$=$(i_{la}, i_{lb}, i_{lc})^T$ are the load currents,
while $i$=$(i_a, i_b, i_c)^T$ are the filter currents. $L$ indicates the value of the inductances, and $C$ the DC-link bus capacitor value.

\subsection{Mathematical model}
\begin{figure}[t]	
\centering
\includegraphics[width=5cm,height=5cm]{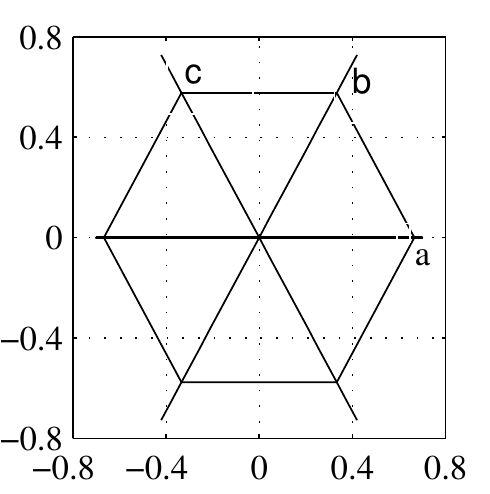} 
\caption{Hexagon of feasible $u_{abc}$.}\label{fig:esagono_u}
\end{figure}

Considering the inductors dynamics, the filter model can be expressed as
\begin{equation}\label{eq:modelabc}
\begin{gathered}
\begin{bmatrix}
v_{ma}(t) \\
v_{mb}(t) \\
v_{mc}(t)
\end{bmatrix}-L\frac{d}{dt} \begin{bmatrix}
i_a(t)\\
i_b(t)\\
i_c(t)
\end{bmatrix}-R\begin{bmatrix}
i_a(t)\\
i_b(t)\\
i_c(t)
\end{bmatrix}
=\begin{bmatrix}
u_x(t)\\
u_y(t) \\
u_z(t)
\end{bmatrix}v(t)-v_{NK}\begin{bmatrix}
1 \\
1 \\
1
\end{bmatrix}
\end{gathered}
\end{equation}
where $R$ is the parasitic resistance related to the inductance $L$ and to the cables, $v_{NK}$ is the voltage
between the nodes $N$ and $K$ reported in Fig. \ref{fig:SAFscheme}, $v(t)$ is the voltage on the DC-link capacitor, and $u_1=(u_x,u_y,u_z)^T$ is the  switch command vector for the legs of the converter. Since a PWM (Pulse Width Modulation) strategy is assumed to control the inverter, the above-mentioned control inputs can be considered such that $u_{1i}\in[0, \ 1], \ \ i=x,y,z$. According to the three-wire topology for any generic voltage/current vector $x$ it holds
\begin{equation}
\sum_{i=a,b,c}{x_i}=0
\end{equation}
hence, from the sum of the scalar equations in (\ref{eq:modelabc}) it follows that
\begin{equation}\label{eq:vincolo}
v_{NK}=\frac{u_x(t)+u_y(t)+u_z(t)}{3}v(t)
\end{equation}
defining
\begin{equation}
\begin{gathered}
u_{abc}=[u_a(t),\ u_b(t),\ u_c(t)]^T=\begin{bmatrix}
u_x(t)\\
u_y(t)\\
u_z(t)
\end{bmatrix}-\frac{u_x(t)+u_y(t)+u_z(t)}{3}\begin{bmatrix}
1\\
1\\
1
\end{bmatrix}
\end{gathered}
\end{equation}
it can be verified by direct computations that
\begin{equation}\label{eq:vincolo1}
[1\ 1\ 1]u_{abc}(t)=0 \ \ \forall t \geq 0.
\end{equation}
For what concerns the state equation relative to the capacitor voltage dynamics, it can be derived considering an ideal inverter
and applying a power balance condition between the input and the output of the filter, then replacing (\ref{eq:vincolo}) into (\ref{eq:modelabc}), the complete filter model results
\begin{equation}\label{eq:modabc_comp}
\begin{aligned}
\frac{di}{dt}&=-\frac{R}{L}I_3i(t)-\frac{v(t)}{L}u_{abc}(t)+\frac{1}{L}v_{mabc}\\
\frac{dv}{dt}&=\frac{1}{C}u_{abc}^T(t)i(t)
\end{aligned}
\end{equation}
where the filter currents dynamics have been written in a more compact form with respect to (\ref{eq:modelabc}), multiplying the current vector by the identity matrix of suitable dimension $I_3$.
Exploiting equations (\ref{eq:vincolo}), (\ref{eq:vincolo1}), the system model
can be reduced to the standard \emph{two-phase planar representation} of a three-phase balanced systems \citep{Krause95}, which can be obtained applying the following coordinates transformation
\begin{equation}
\begin{aligned}
i_{\alpha \beta}(t) &=[i_{\alpha} \ i_{\beta}]^T= ^{\alpha  \beta}T_{abc}i(t) \\
\ u_{\alpha \beta}(t)&=[u_{\alpha} \ u_{\beta}]^T= ^{\alpha \beta}T_{abc}u_{abc}(t) \\
v_{m\alpha \beta}&=[v_{m\alpha} v_{m\beta}]^T= ^{\alpha \beta}T_{abc}v_m\\
^{\alpha \beta}T_{abc}&=\frac{2}{3}\begin{bmatrix}
1 & -\frac{1}{2} & -\frac{1}{2} \\
0 & \frac{\sqrt{3}}{2} & -\frac{\sqrt{3}}{2}
\end{bmatrix}
\end{aligned}
\end{equation}
the SAF dynamics expressed in this $\alpha-\beta$ reference frame become
\begin{equation}\label{eq:modalfabeta}
\begin{aligned}
\frac{di_{\alpha \beta}}{dt}&=-\frac{R}{L}I_2i_{\alpha \beta}(t)-\frac{v(t)}{L}u_{\alpha \beta}(t)+\frac{1}{L}v_{m \alpha \beta}\\
\frac{dv}{dt}&=\frac{3}{2C}u_{\alpha \beta}^T(t)i_{\alpha \beta}(t)
\end{aligned}
\end{equation}
according to the hypothesis of three-phase balanced sinusoidal line, the ideal main voltage tern can be expressed in the above-defined bi-dimensional reference frame as follows
\begin{equation*}
[v_{m\alpha} \ v_{m \beta}]^T=V_m[cos(\omega_m) \ sin(\omega_m)]^T
\end{equation*}
where $V_m$ is the voltage amplitude and $\omega_m$ the grid angular frequency.
For what concerns the control vector $u_{abc}$, in this reference frame the eight possible configurations of the switching network (reported in Tab. \ref{tab:switch}) can be mapped in the $\alpha-\beta$ plane, obtaining the vertexes and the origin of the feasibility space  illustrated in Fig.\ref{fig:esagono_u}, while each point in the hexagon can be obtained as mean value in a PWM period.
\begin{table}[t]
\centering
\begin{tabular}{|c c c|c c c | c c|} \hline
$u_x$ & $u_y$ & $u_z$ & $u_a$ & $u_b$ & $u_c$ & $u_{\alpha}$ & $u_{\beta}$ \\  \hline
0 & 0 & 0 & 0 & 0 & 0 & 0 & 0   \\ \hline
1 & 0 & 0 & 2/3 & -1/3 & -1/3 & 2/3 & 0   \\ \hline
1 & 1 & 0 & 1/3 & 1/3 & -2/3 & 1/3 & 1/$\sqrt{3}$   \\ \hline
0 & 1 & 0 & -1/3 & 2/3 & -1/3 & -1/3 & 1/$\sqrt{3}$  \\ \hline
0 & 1 & 1 & -2/3 & 1/3 & 1/3 & -2/3 & 0  \\ \hline
0 & 0 & 1 & -1/3 & -1/3 & 2/3 & -1/3 & -1/$\sqrt{3}$  \\ \hline
1 & 0 & 1 & 1/3 & -2/3 & 1/3 & 1/3 & -1/$\sqrt{3}$  \\ \hline
1 & 1 & 1 & 0 & 0 & 0 & 0 & 0  \\ \hline
\end{tabular}
\caption{Control function table.}\label{tab:switch}
\end{table}
As it will become clear in the next sections, in order to simplify the control objectives definition and the controller design, it is very useful to adopt a further transformation from the two-phase current variables $[i_{\alpha} \ i_{\beta}]^T$ to a two-phase real-virtual (imaginary) power variables defined as
\begin{equation}
x=[x_d \ x_q]^T= ^{dq}T_{\alpha \beta}i_{\alpha \beta}
\end{equation}
where
\begin{equation*}
^{dq}T_{\alpha \beta}=V_m\begin{bmatrix}
cos(\omega_m t) & sin(\omega_m t) \\
-sin(\omega_m t) & cos(\omega_m t)
\end{bmatrix}.
\end{equation*}
In this so-called \emph{synchronous} coordinate setting, aligned with the mains voltage vector, the model of the SAF is expressed as
\begin{equation}\label{eq:modeldq}
\begin{aligned}
\dot{x}&=M(R,L)x-\frac{v}{L}u_{dq}+d_0\\
\dot{v}&=\frac{\epsilon}{2}u_{dq}^T x
\end{aligned}
\end{equation}
where
\begin{equation}\label{eq:definitions}
\begin{gathered}
d_0=\begin{bmatrix}
E_{md}/L \\
0
\end{bmatrix}, \ M(R,L)=\begin{bmatrix}
-R/L & \omega_m \\
-\omega_m & -R/L\end{bmatrix},
\ \epsilon=\frac{3}{CE_{md}}, \ E_{md}=V_m^2, \ u_{dq}= ^{dq}T_{\alpha \beta}u_{\alpha \beta}
\end{gathered}
\end{equation}
it is further to notice that, since the filter currents, the mains voltage and the DC-link voltage are measurable,
the full state ($x$,$v$) is available for feedback, moreover the actual control action $u=[u_x \ u_y \ u_z]$ can be
determined from $u_{abc}$, which in turn can be derived from $u_{dq}$.\\
As regards the load description, the same two-phase real-virtual power representation can be used, in particular following \citep{Akagi1984}, the load currents can be approximated as periodic signals given by the sum of a finite number N of harmonics, with frequencies multiple of $f_m=\omega_m/2\pi$. Hence the load currents can be expressed in power variables as
\begin{equation}\label{eq:loaddq}
x_{lj}=X_{lj0}+\sum_{n=1}^{N+1}X_{ljn}cos(n\omega_mt+\psi_{jn}), \ \ j=d,q
\end{equation}
where the harmonics amplitudes $X_{ld0}$, $X_{lq0}$, $X_{ldn}$, $X_{lqn}$ and phases $\psi_{dn}$,  $\psi_{qn}$ are constants.
Since the load currents and the mains voltages are measurable, also the variables ($x_{ld}$, \ $x_{lq}$) will be considered known and available for control purpose.
\subsection{Problem statement and control objectives}\label{sec:prob_stat}
Roughly speaking the main control objective of the considered SAF is to steer the  variables $x_d$, $x_q$, injecting power into the line to compensate for the load harmonics. However the ability of tracking current references relies upon the energy stored in the DC-link capacitor, which is the main power source of the filter, therefore another general objective is to keep the DC-link voltage confined in a suitable region, to avoid overcharge and, at the same time, to ensure the capability to steer the filter currents. On the other hand the ability of maintaining DC-link voltage into a suitable region is strictly related to the power exchanged with the mains, which in turn is affected by the current harmonics to be compensated for. The general control objective is then two-folds; one related to the tracking of current disturbances, the other concerns the voltage internal dynamics stabilization. In this paragraph a precise and feasible control problem is formally defined, recalling the considerations made above, and assuming that a suitable dimensioning, that will be deeply discussed in the next section, has been carried out.\\
Bearing in mind the power variables representation of a generic nonlinear load expressed in (\ref{eq:loaddq}), it turns out that the only desired load component is $X_{ld0}$, since it represents first-order harmonics aligned with the mains voltages, while the remaining part of the real component $x_{ld}-X_{ld0}$ is an oscillatory signal with null balance over a line period, and the imaginary component $x_{lq}$ represent a measure of the misalignment between mains ideal voltage and load currents (see \cite{Mohan1989}) and do not contribute to the power flow. In this respect, the terms $x_{ld}-X_{ld0}$, $x_{lq}$ are undesired components which should be canceled by the injected filter currents, hence ideally the control problem can be formulated as a state tracking problem, for system (\ref{eq:modeldq}), of the following reference
\begin{equation}\label{eq:idealref}
x^{*}(t)=[x^{*}_{d} \ \ x^{*}_{q}]^T=\left[X_{ld0}-x_{ld} \ \ -x_{lq} \right]^T
\end{equation}
a prefect tracking of this reference would ensure pure sinusoidal mains currents perfectly aligned with the mains voltages. However, this ideal objective is in contrast with the requirement to have a DC-link voltage bounded behavior. In order to formally motivate this claim, consider the steady state voltage dynamics in case perfect tracking of the power reference $x^{*}(t)$ is achieved, after some computations it results
\begin{equation}
\frac{dv^2}{dt}=\epsilon L\left(d_0+M(R,L)x^{*}(t)-\dot{x}^*(t) \right)^Tx^*:=\epsilon \Psi(x^*(t))
\end{equation}
the signal $\epsilon \Psi(x^*(t))$ which drives the integrator is periodic with period $T=1/f_m$, and it is composed by the sum of a zero mean value signal $\epsilon L (d_0-\dot{x}^*)^Tx^*$, and the signal $\epsilon L (M(R,L){x}^*)^Tx^*$ which has negative mean value as long as parasitic resistance $R$ or reference $x^*$ are not zero. By this, no matter the starting voltage value of the DC-link, the capacitor will be discharged and the voltage will drop, providing a loss of controllability of the system.\\
To avoid this phenomenon, the reference must be revised, taking into account an additional power term, which should be drained from the line grid  by the active filter, in order to compensate for its power losses. Following this motivation, and recalling that the unique useful component for the energy exchange is the real part of the power variables, the ideal reference signal (\ref{eq:idealref}) is modified as
\begin{equation}\label{eq:refreal}
x^*_{\varphi_0}=x^*+(\varphi_0 \ \ 0)^T
\end{equation}
in which $\varphi_0$ is a solution of the following equation
\begin{equation}\label{eq:powbal}
R\varphi_0^2-E_{md}\varphi_0+Rf_m \int_{0}^{1/f_m} (x_d^{*2}(\tau)+x_q^{*2}(\tau))d\tau=0
\end{equation}
this represents the power balancing condition which guarantees that the internal voltage dynamics in case of perfect tracking of the modified reference $x_{\varphi_0}^*$ is
\begin{equation}
\frac{dv^2(t)}{dt}=\epsilon \Psi(x_{\varphi_0}^*(t))
\end{equation}
and the right hand side $\epsilon \Psi(x_{\varphi_0}^*)$ is periodic with period $1/f_m$ with zero mean value. A brief discussion is needed for the solutions of equation (\ref{eq:powbal}), it has two real positive solutions if the following condition is verified
\begin{equation}\label{eq:condreal}
E_{md}^2\geq 4R^2f_m\int_{0}^{1/f_m}(x_d^{*2}(\tau)+x_q^{*2}(\tau))d\tau
\end{equation}
from a physical viewpoint relation (\ref{eq:condreal}) set an upper bound on the admissible undesired components which can be compensated and on the parasitic resistance $R$, however, as typically $E_{md}>>R$, this condition is not limitative at all.
The two solutions of (\ref{eq:powbal}) under condition (\ref{eq:condreal}) are
\begin{equation}
\begin{aligned}
\varphi_0 & \approx \frac{R}{E_{md}}f_m\int_{0}^{1/f_m}(x_d^{*2}(\tau)+x_q^{*2}(\tau))d\tau \approx 0 \\
\varphi_0 & \approx \frac{E_{md}}{R}
\end{aligned}
\end{equation}
the first solution, minimizing the power drained from the line grid to compensate the power losses, is the physically most plausible, because the power consumed by parasitic resistances in the filter is usually quite small, hence it will be considered throughout the chapter.\\
The control problem which will be faced in this work can now be precisely stated; the issue is to design the control vector
$u_{abc}$ in a way such that the following objectives are fulfilled:
\\
$A$) Given the reference signal $x_{\varphi_0}^*$ defined in (\ref{eq:refreal}), asymptotic tracking must be achieved, that is
\begin{equation}
\lim_{t\rightarrow \infty} (x(t)-x^*_{\varphi_0})=0;
\end{equation}
$B$) Given a safe voltage range $[v_m, \ v_M]$, with $v_M>v_m>0$, and assuming $v(t_0)\in[v_m, \ v_M]$, it is required that
\begin{equation}
v(t)\in[v_m, \ v_M], \forall t>t_0;
\end{equation}
\\
it can be verified that the tracking of the modified power reference is potentially achievable keeping the voltage dynamics inside the safe region, only if the zero mean value oscillating component of $\epsilon\Psi(x^*_{\varphi_0})$ is properly bounded, this can be ensured by a suitable capacitor design.\\
In the regulator  design, saturation of the actual input $u_1$, imposed by PWM strategy, will not be taken explicitly into account, also this approximation takes advantage of a correct sizing methodology; as it will become clear in Section \ref{sec:dimensionamento}, a suitable choice of the DC-link voltage lower bound $v_m$, depending on the currents to be compensated for, has to be made to meet the constraint $u_1\in[0, \ 1]$, at least when the power tracking error is reasonably small.\\
A further consideration needs to be made on the requirement $v(t_0)\in[v_m, \ v_M]$; according to the AC/DC boost converter theory \citep{Mohan1989}, the natural response of the system would lead the DC-link voltage at twice the line voltage peak value, due to the resonant behavior of the $LC$ pair and the free-wheeling diodes of the switching bridge. If a proper design has been performed, this value is expected to be greater than the voltage range lower bound $v_m$; hence, after a transient period, the controller can be switched on having the initial voltage value inside the admissible region as required by objective $B$.\\
Finally it is further to remark that $x^*_{\varphi_0}$ depends on parasitic resistance $R$ through (\ref{eq:powbal}), hence it has to be considered as an unknown variable, to be reconstructed by estimating the power losses by means of a proper elaboration of the DC-link voltage signal.%
\section{Shunt active filter sizing methodology} \label{sec:dimensionamento}
The aim of this section is to present a precise algorithm to properly select the SAF hardware components, two different design objectives are considered, the first is to select the minimal component values dependent on the level of current distortion imposed by the load, while the second is to find the minimum capacitor value necessary to compensate all the possible loads compatible with the maximum current rating of the inverter switching devices. Both the methods are control-oriented, that is they ensure the feasibility of control objectives stated in \ref{sec:prob_stat} and that control input saturation is avoided under nominal load and line voltage conditions.\\
The proposed design method is based on the model derived in Section \ref{sec:modello}, a further approximation is considered with respect to equation (\ref{eq:modabc_comp}); the inductors are modeled as pure inductance, that is the parasitic resistance $R$ is neglected, while ideal mains voltage tern and converter switches are considered as in the previous section.
\subsection{Inductance value selection}
\begin{figure}[t]	
\centering
\psfrag{p1}{\begin{small}$u_1$ \end{small}}
\psfrag{p2}{\begin{small}$u_2$ \end{small}}
\psfrag{p3}{\begin{small} $u^*$ \end{small}}
\includegraphics[width=4cm,height=4cm]{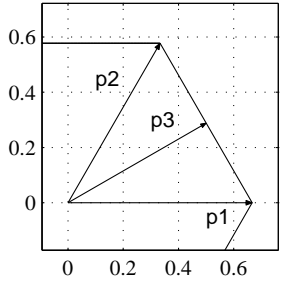} 
\caption{Current ripple worst case.}\label{fig:IrippleWC}
\end{figure}

The inductance value can be selected regardless the loads, hence this part of the design procedure is the same for both the design objectives previously defined.\\
The design criterion is based on the maximum current ripple $\Delta I_{Mpp}$ allowed for the filter currents; current ripple is a consequence of the PWM technique applied to obtain the reference command value $u_{abc}^{*}$, it has to be bounded in order to limit high frequency distortion. The actual command vector $u_{abc}(t)$ and filter current $i(t)$ are affected by a ripple component
\begin{equation}\vspace{1mm}
\begin{aligned}
i(t)&=i^*(t)+\Delta i(t) \\[2mm]
u_{abc}(t)&=u_{abc}^*(t)+\Delta u_{abc}(t)
\end{aligned}\vspace{1mm}
\end{equation}
substituting these expressions in the state equation (\ref{eq:modabc_comp}) it turns out
\begin{equation}\vspace{1mm}
L\frac{\Delta i(t)}{dt}=-\Delta u_{abc}(t)v(t)\vspace{1mm}
\end{equation}
 by simple computation it can be showed that the worst ripple case occurs when the desired command value $u_{abc}^*$ is in the middle of a feasibility hexagon side (see Fig. \ref{fig:IrippleWC}). In this condition, assuming that the DC-link voltage has constant value $V$ in a PWM period, the peak to peak current ripple is
\begin{equation}\vspace{1mm}
\Delta I_{pp}=\int_{t}^{t+T_s/2}\frac{d \Delta i(t)}{dt}dt=\frac{V}{6f_{PWM}L}\vspace{1mm}
\end{equation}
where the sampling period $T_s$ and the PWM frequency $f_{PWM}$ are assumed already set before starting the sizing procedure. If the peak to peak ripple must be bounded by the
desired maximum value $\Delta I_{Mpp}$, the following inequality needs to be fulfilled
\begin{equation}\label{eq:dimL}
L \geq \frac{v_M}{6f_{PWM}\Delta I_{Mpp}}\Rightarrow L_{min}=\frac{v_M}{6f_{PWM}\Delta I_{Mpp}}
\end{equation}
the upper bound of the voltage range $v_{M}$ depends only on the kind of capacitor and it can be supposed already chosen before starting the design procedure, hence the minimum inductance value $L_{min}$ compatible with the desired maximum current ripple can be selected applying equation (\ref{eq:dimL}).
\subsection{Load-based approach}
\begin{figure}[t]	
\centering
\includegraphics[width=4.3cm,height=4.3cm]{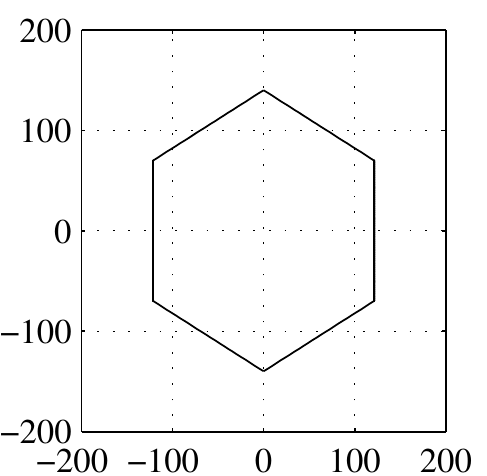} 
\caption{Hexagon of feasible filter current.}\label{fig:esagonoI}
\end{figure}

Let us now consider the first design algorithm based on the knowledge of the load to be compensated for. The load distortion will be modeled as in equation (\ref{eq:loaddq}), taking into account the constraint on the maximum current $I_{max}$ of the device implementing the bridge switches. The switching devices sizing depends on the total amount of power (distorted and reactive) $P=3V_{mRMS}I_{SAFRMS}$ that the filter has to compensate for (if the load is known then $P$ is known), hence by the route mean square value $I_{SAFRMS}$, the maximum current that the switches  need to drain can be readily obtained as $I_{max}=\sqrt{2}I_{SAFRMS}$.\\
The desired filter currents (denoted with $^*$) necessary to fulfill the tracking objective $A$ defined in \ref{sec:prob_stat} can be effectively imposed by the converter if each component is less than the maximum allowed value, i.e
\begin{equation}\label{eq:vincoloImax}
i^*(t)=[i_a^*(t) \ i_b^*(t) \  i_c^*(t)]^T\leq I_{max}[1 \ 1 \ 1]^T, \ \forall t
\end{equation}
this feasibility condition can be graphically represented considering that each projection of the filter currents vector must be less then $I_{max}$, hence the feasibility space is an hexagon similar to that reported in Fig. \ref{fig:esagonoI} (obtained taking $P=45kVAR$ as filter size, $V_{mRMS}=220V$ and then $I_{max}=70A$). Therefore condition (\ref{eq:vincoloImax}) can be readily checked considering the inscribed circle in the feasibility hexagon.
If the load currents do not satisfy constraint (\ref{eq:vincoloImax}) the number of current harmonics to be compensated for has to be reduced, differently, when the filter performance cannot be decreased, the opportunity to connect two shunt active filters to the same load can be considered.\\
Assuming that an inductance value such that $L \geq L_{min}$ has been selected, the voltages at the input of the six switches bridge can be calculated as
\begin{equation}\label{eq:vqd_load}
\begin{gathered}
v_{dq}^{*}(t)=v(t)u_{dq}^*(t)=\begin{bmatrix}
V_{m} \\
0
\end{bmatrix}-L\frac{di^*_{dq}}{dt}+\begin{bmatrix}
0 & \omega_m \\
-\omega_m & 0
\end{bmatrix}i_{dq}^*
\end{gathered}
\end{equation}
the above equation is obtained by inversion of equation (\ref{eq:modeldq}) with $R=0$ and expressing the model in the \emph{synchronous} reference frame in current rather than in power variables, in order to directly consider the load currents in the design approach.\\
The constraints on the command inputs need to be considered too, by (\ref{eq:vqd_load}) the inductance value must be as low as possible in order to make $u_{dq}^*$ feasible, taking into account also the current ripple limitation we select $L=L_{min}$. As mentioned, the choice of the of the capacitor voltage lower bound value  plays a key role to avoid saturation issues on command inputs, this can be easily verified approximating the hexagon in Fig. \ref{fig:esagono_u} with the inscribed circle. In order to avoid control action saturation (assuming perfect power tracking) it must be imposed that
\begin{equation}\label{eq:dimvm}\vspace{1mm}
||u_{abc}^*||=\frac{||v_{abc}^*(t)||}{v(t)}\leq \frac{||v_{abc}^*(t)||}{v_{m}} \leq r_{in}=\frac{1}{\sqrt{3}}, \ \ \forall t \vspace{1mm}
\end{equation}
with $r_{in}$ the radius of the inscribed circle and $v_{abc}^*=v(t)u_{abc}^*$. From (\ref{eq:dimvm}) design equation for $v_{m}$ can be obtained
\begin{equation}\label{eq:vmdis}\vspace{1mm}
v_{M}\geq v_{m} \geq \frac{||v_{abc}^*(t)||}{r_{in}}, \forall t \in \bigg[\frac{n}{f_m},\frac{n+1}{f_m}\bigg], \ n=0,1,\dots \vspace{1mm}
\end{equation}
usually $v_m$ is oversized with respect the value given by the inequality above, in order to avoid saturation even if non-zero tracking errors are present.
If condition (\ref{eq:vmdis}) cannot be satisfied, some alternatives need to be considered; the capacitor can be changed in order to adopt an higher upper bound $v_{M}$, when the costs of the project have to be limited and the kind of capacitor cannot be substituted, the number of harmonics considered must be reduced until (\ref{eq:vmdis}) is satisfied. To preserve the number of harmonics to compensate, the inductance value can be reduced, penalizing the current ripple and then tolerating a greater high frequency distortion.\\
The capacitor value can then be selected assuming an ideal converter and writing the balance equation between the instantaneous reference power at the input of the six switches bridge and the power of the DC-link capacitor, hence
\begin{equation}\vspace{1mm}
p_{filt}(t)=[v_{dq}(t)]^Ti_{dq}^*(t)=\frac{d}{dt}\left(\frac{1}{2}Cv^2(t) \right)
\end{equation}
the corresponding energy can be calculated as
\begin{equation}
E_{filt}(t)=\int_{t_0}^t p_{filt}(\tau)d\tau
\end{equation}
by the hypothesis of sinusoidal load currents and ideal mains voltages $E_{filt}(t)$ is periodic of frequency $f_m$ and its mean value is zero. Defining
\begin{equation}
\begin{aligned}
E_{max}&=max|E_{filt}(t)| \\
v_{ref}&=\frac{v_{M}+v_{m}}{2}
\end{aligned}
\end{equation}
and imposing that the voltage variation corresponding to $E_{max}$ is $v_{ref}-v_{m}$, the capacitor value design equation can be written as
\begin{equation}\label{eq:dimC}
C=\frac{2E_{max}}{v_{ref}^2-v_{m}^2}
\end{equation}
\subsection{Switches-based approach}
As stated at the beginning of this section, a different design method aims to find the capacitor value that makes the filter able to compensate for the worst load compatible with the switches maximum current. If the resulting capacitor value is not too expensive, this method allows to design the filter only knowing the amount of current that has to be compensated.\\
During the optimization procedure the load currents need to be the only varying parameters while all the other values must be fixed. The inductance value is chosen equal to the minimum compatible with the allowed ripple, while the minimum capacitor voltage $v_{m}$ is supposed sufficiently low to make simple the voltage control, and, at the same time, the resulting capacitor value  feasible.
Writing the filter currents spectrum in the $d-q$ synchronous reference frame, an expression similar to (\ref{eq:loaddq}) can be obtained
\begin{equation}
i_{j}(t)=I_{j0}+\sum_{n=1}^{N+1}I_{jn}cos(2\pi nf_m t+\psi_{jn}), \ \ j=d,q
\end{equation}
the parameters to be varied in order to calculate the worst $E_{max}$ are the $(2N+1)+1$ magnitudes and the $2N+1$ phases, so the following optimization problem
\begin{equation}
E_{max}^{worst}=\max_{z}\max_{t} |\int_{t_0}^t [v_{dq}(\tau)]^Ti_{dq}(\tau)d\tau|
\end{equation}
has to be solved with respect to the array $z$ of $4(N+1)+1$ variables, taking into account the following  constraints
\begin{itemize}
  \item switches currents must be less than the maximum allowed, that is the current vector must be inside an hexagon similar to that reported in Fig. \ref{fig:esagonoI}. This can be easily checked approximating the hexagon with its inscribed circle;
  \item the control output must be feasible, that is the vector $u_{abc}$ must be inside the hexagon reported in Fig. \ref{fig:esagono_u}. This can be easily checked approximating the hexagon with its inscribed circle;
  \item harmonics components phases have to be greater than $-\pi$ and less than $\pi$.
\end{itemize}

Once $E_{max}^{worst}$ has been determined, substituting its value in (\ref{eq:dimC}), the capacitor value relative to the switches-based design approach can be selected.\\
In the discussion above, ideal mains voltages have been assumed, if also the grid line voltages are distorted, the capacitor has to provide more energy to the load, hence its value must be higher than the one calculated under ideal conditions.\\
In case of ideal mains voltages the load instant power is the one calculated in (\ref{eq:loaddq}) and the only power term that the filter must deliver is $x_{ldn}=\sum_{n=1}^{N+1}X_{ldn}cos(n\omega_m t+\psi_{dn})=V_m i_{ldn}$ having zero mean value. If the mains voltages are distorted, their representation in the synchronous reference frame is
\begin{equation}\vspace{1mm}
v_{mdq}(t)=[V_m+v_{mdn}, \ v_{mq}]^T\vspace{1mm}
\end{equation}
line voltages harmonic perturbation produces additive terms in the load instantaneous power expression, that by direct computation can be written as
\begin{equation}\vspace{1mm}
p_{ladd}=v_{mdn}(t)i_{ldn}(t)+v_{mq}i_{lq}(t)\vspace{1mm}
\end{equation}
the above equation shows that the filter has to provide more power to the load, furthermore the power mean value in a PWM period can be different from zero. Hence also assuming that the mean value becomes zero in a finite time, the capacitor must be oversized with respect to the ideal situation, in order to accumulate more energy.
\section{Robust controller design} \label{sec:controllo}
\begin{figure}[t]	
\centering
\includegraphics[width=6cm, angle=-90]{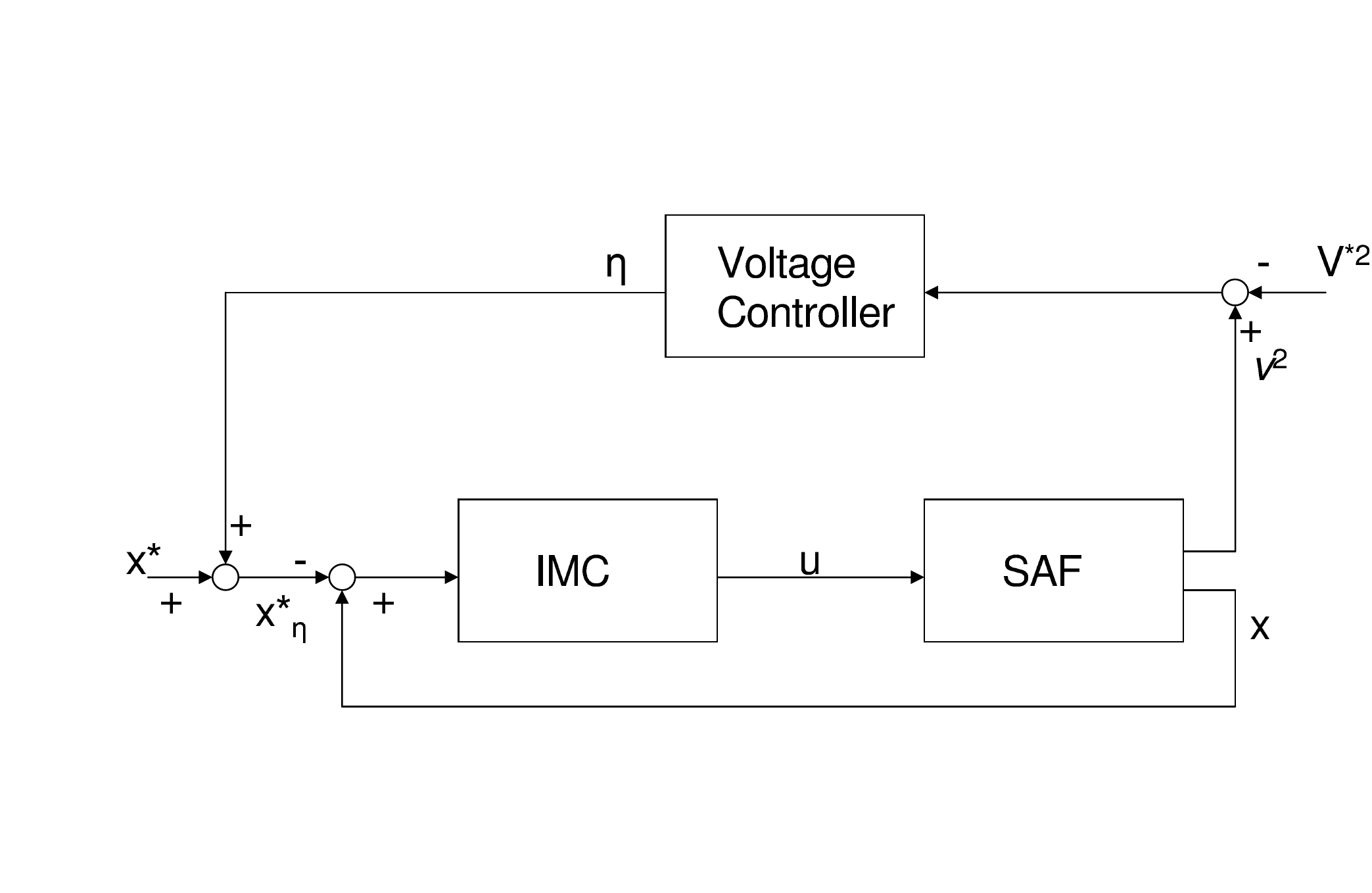} 
\caption{Controller structure.}\label{fig:controlblock}
\end{figure}

In this section the control problem defined in \ref{sec:prob_stat} is addressed, relying upon a suitable capacitor value given by the procedure described in the previous section, the two interlaced objectives $A$ and $B$ defined in \ref{sec:prob_stat} can be approached individually by exploiting the principle of singular perturbation. Two independent controllers (reported in the block diagram of Fig. \ref{fig:controlblock}) will be designed. An internal model-based controller (IMC) is proposed in order to deal with the problem of robust reference tracking (defined in objective $A$) for the fast subsystems composed by the power variables dynamics, while an independent voltage controller for the slow DC-link voltage subsystem is designed to produce a reference modification $\eta$ which compensate the unknown power losses term $\varphi_0$, allowing to achieve objective $B$. The averaged voltage value is chosen as the controlled variable, and a phasor variables representation is exploited to design the regulator, this choice allows for the necessary voltage oscillation during nominal operation, and improves the voltage dynamics behavior with respect to other proposed solutions \citep{Marconi2007}. Stability analysis is carried out in two steps; the \emph{reduced averaged dynamics}, obtained replacing the steady state of the fast subsystem into the slow voltage dynamics and carrying out the average value to obtain a phasor variables representation, and the \emph{boundary layer system}, obtained considering the SAF currents dynamics and an ideal energy storage element, are proved to be asymptotically stabilized by the proposed controllers. Then practical stability for the overall closed-loop error system is stated exploiting well-established singular perturbation and two time-scale systems theory results.\\
Before detailing the proposed control structure, consider the first preliminary control law
\begin{equation}\label{eq:utens}\vspace{1mm}
\bar{u}(t)=v(t)u_{dq}(t)
\end{equation}
which is always well defined provided that $v(t)\geq v_{m}>0$ for all $t \geq 0$ according to objective $B$.
Replacing (\ref{eq:utens}) into (\ref{eq:modeldq}) yields
\begin{equation}\label{eq:modelcontr}
\begin{aligned}
\dot{x}&=M(R,L)x+\frac{1}{L}\bar{u}+d_0 \\
\frac{dv^2}{dt}&=\epsilon\bar{u}^Tx
\end{aligned}
\end{equation}
now consider the modified power reference
\begin{equation}\label{eq:refeta}
x_{\eta}^*=x^{*}+(\eta \ \ 0)^T
\end{equation}
and define the change of variables
\begin{equation}\label{eq:errvar}
\tilde{x}=x-x_{\eta}^*, \quad \tilde{z}=v^2-V^{*2}
\end{equation}
where $V^{*2}=(v_m^2+v_{M}^2)/(2)$ is the reference value for the square DC-link voltage. Note that the requirement B of having $v(t)\in [v_m \ \ v_{M}]$ for all $t \geq t_0$ can be equivalently formulated in the error variable $\tilde{z}$ requiring $\tilde{z}(t) \in [-l^* \ \ l^*]$ for all $t \geq t_0$, with $l^*=(v_M^2-v_m^2)/2$. The complete system (\ref{eq:modelcontr}) can be then expressed in the error variables defined in (\ref{eq:errvar}), the transformed model results
\begin{equation}\label{eq:errdyn}
\begin{aligned}
\dot{\tilde{x}}&=M(R,L)\tilde{x}-\frac{1}{L}\bar{u}+d_0-\dot{x}_{\eta}^*+M x_{\eta}^*\\
\dot{\tilde{z}}&=\epsilon \bar{u}^T[\tilde{x}+x_{\eta}^*].
\end{aligned}
\end{equation}
The controller design will be carried out considering the error dynamics in (\ref{eq:errdyn}), in summary the idea is to steer the closed loop dynamics toward a steady state in which $\tilde{z}$ is free to oscillate within the admissible region, but its mean value is steered to zero (i.e the DC-link voltage mean value is steered to $V^*$), and $\tilde{x}$ is steered to zero, i.e the power $x$ follows a reference which is the sum of the term $x^*$, which takes into account the undesired harmonic load components, and a constant bias $\eta$ which is needed in order to compensate the power losses and to make the range $[v_m \ \ v_{M}]$ an invariant subspace for the voltage dynamics.
\subsection{Averaging voltage controller}\label{sec:AveragingV}
In order to fulfill objective $B$ the voltage dynamics need to be stabilized, in this respect the subsystem composed by the capacitor voltage dynamics  will be considered, a suitable \emph{reduced averaged system} will be sought, and then a controller for the capacitor voltage DC component will be designed.\\
The first step is to average the voltage differential equation to obtain the dynamics in the so-called \emph{phasor-variables}, then, a control law, itself expressed on phasor representation, can be designed following an approach similar to that proposed in \citep{Valderrama2001}, however in this work the only voltage subsystem is controlled using phasor variables, while the power subsystem is controlled in the real time domain.\\
The controlled variable is chosen to be the time-window averaged value $\tilde{z}_a$ of the square voltage error $\tilde{z}$, and the averaging is performed over the time interval $[t-T,t]$. In terms of \citep{Sanders1991} this average value is a zero-order phasor defined as
\begin{equation}\label{eq:vaveraged}
\tilde{z}_a(t)=\int_{t-T}^{t}\tilde{z}(\tau)d\tau
\end{equation}
the fact that $\tilde{z}_{a}$ is a zero-order phasor allows to obtain its derivative by simply applying the same averaging procedure to its differential equation in (\ref{eq:errdyn})
\begin{equation}\label{eq:averageder}
\dot{\tilde{z}}_a=\frac{1}{T}\int_{t-T}^{t}\tilde{z}(\tau)d\tau=\epsilon \int_{t-T}^{t} \bar{u}^T[\tilde{x}+x_{\eta}^*]d\tau
\end{equation}
note that the average voltage derivative can also be expressed as the difference over one period of the actual voltage, hence
\begin{equation}\label{eq:dotza}
\frac{d}{dt}(\tilde{z}_a)=\frac{d}{dt}\int_{t-T}^{t}\tilde{z}(\tau)d\tau=\frac{\tilde{z}(t)-\tilde{z}(t-T)}{T}
\end{equation}
this insight connotes the availability of $\tilde{z}_a$ for measurement in real time, and, as it will later clarified, it is of crucial importance for an actual implementation of the controller.\\
All further elaborations will focus on the integral-differential equation (\ref{eq:averageder}) representing the averaged error voltage dynamics. This equation depends on $\bar{u}$ which is actually provided by the power tracking controller, to eliminate $\bar{u}$ consider that the differential equation for $\tilde{x}$ in (\ref{eq:errdyn}) can be rewritten as
\begin{equation}\label{eq:nou}
\bar{u}=L(M(R,L)x_{\eta}^*-\dot{x}_{\eta}^*+M(R,L)\tilde{x}+d_0-\dot{\tilde{x}})
\end{equation}
replacing (\ref{eq:nou}) into (\ref{eq:averageder}) the following equation is obtained
\begin{equation}\label{eq:zanou}
\dot{\tilde{z}}_a=\frac{\epsilon L}{T}\int_{t-T}^t (M(R,L)x_{\eta}^*-\dot{x}_{\eta}^*+d_0)^Tx_{\eta}^*d\tau+\epsilon L\tilde{D}(\tilde{x})
\end{equation}
where $\tilde{D}(\tilde{x})$ collects all the terms depending on the power tracking error $\tilde{x}$. The next step is to exploit the
fact that the reference term $x^*$ is $T$-periodic ($T=1/f_m$), hence it results in a constant value when averaged over this period, this is a key advantage of the averaging approach for the voltage system. The $T$-periodic terms in (\ref{eq:zanou}) can be summarized to
\begin{equation}
D^*=\frac{1}{T}\int_{t-T}^t[(M(R,L)x^*-\dot{x}^*+d_0)^Tx^*]d\tau
\end{equation}
since $x^*$ is periodic in $T$, $D^*$ is a constant disturbance, and, due to power losses induced by  the parasitic resistance $R$, it also follows that $D^*<0$. For further simplification the integral operator can be applied to the occurring derivative terms. Using definitions (\ref{eq:definitions}), (\ref{eq:refeta}), after some computations the averaged error voltage dynamics can be expressed completely in phasor variables
\begin{equation}\label{eq:vdynphasor}
\dot{\tilde{z}}_a=\epsilon[E_{md}\eta_a-2R \nu_a-L\dot{\nu}_a+LD^*+L\tilde{D}]
\end{equation}
where the following nonlinear term has been defined
\begin{equation}\label{eq:defnu}
\nu(t)=\eta(t)\bigg(\frac{1}{2}\eta(t)+x_{d}^*\bigg)
\end{equation}
which enters (\ref{eq:vdynphasor}) with its average and its averaged derivative
\begin{equation}
\begin{aligned}
\nu_a(t)&=\frac{1}{T}\int_{t-T}^t  \nu(\tau)d\tau \\
\dot{\nu}_a(t)&=\frac{\nu_a(t)-\nu_a(t-T)}{T}
\end{aligned}
\end{equation}
the averaged error voltage system is thus controlled by means of the averaged control input
\begin{equation}
\eta_a(t)=\frac{1}{T}\int_{t-T}^t\eta(\tau)d\tau.
\end{equation}
According to singular perturbation theory, the voltage controller design can be carried out considering only the \emph{reduced dynamics}, obtained confusing the value of $\tilde{x}$ with its steady state value $\tilde{x}=0$.\\
As previously remarked, this approximation can be introduced thanks to the small value of $\epsilon$ which, multiplying the voltage dynamics in the second of (\ref{eq:errdyn}), makes the voltage subsystem much slower with respect to the power dynamics in the first of (\ref{eq:errdyn}) (this phenomenon is usually referred as two time-scale system behavior) that will approach the steady state much faster then $\tilde{z}$. Thus reduced voltage dynamics can be obtained by (\ref{eq:vdynphasor}) simply dropping the coupling term $\tilde{D}$, because by definition $\tilde{D}(0)=0$.\\
The nonlinear terms $\nu_a$, and $\dot{\nu}_a$ cannot be managed easily, beside non-linearity they
contain an integral, a time delay and a time-varying term $x_d^*$. In order to simplify the mathematical treatment, a sort of linearized version of system (\ref{eq:vdynphasor}) will be considered. This linear approximation is motivated by several facts; since the parasitic resistance $R$ and the filter inductance value $L$ are usually very small with respect to the term $E_{md}$ in every realistic setup, nonlinear term are much smaller than the linear ones.
Furthermore the component $x_d^*$ has no influence at all in averaging terms if $\eta$ is constant, thanks to the fact that it is $T$-periodic with zero mean value. Hence it will influence the averaged system only while $\eta$ is varying, and also in this case its oscillatory part will be filtered by the averaging procedure. As a result of the previous steps and considerations, the linearized averaged model for the reduced
dynamics can be written as
\begin{equation}\label{eq:averagedlin}
\dot{\tilde{z}}_a=\epsilon E_{md}[\eta_a-\varphi_0]
\end{equation}
where, as mentioned, $\varphi_0$ is the smallest solution of equation (\ref{eq:powbal}).\\
Now it is possible to design the control input $\eta_a$ in order to stabilize the origin of system (\ref{eq:averagedlin}), a standard PI regulator in the averaged variables is proposed
\begin{equation}\label{eq:PIvolt}
\begin{aligned}
\eta_a&=-K_P\tilde{z}_a+\theta \\
\dot{\theta}&=-\epsilon K_I\tilde{z}_a
\end{aligned}
\end{equation}
it is further to notice that the $\epsilon$ factor in the integral action of the controller is introduced
to keep the voltage controller speed in scale with the voltage subsystem to control, thus maintaining the two-time scale behavior of the overall system.\\
In order to prove the asymptotic stability of the closed-loop system resulting by the interconnection of
(\ref{eq:PIvolt}) and (\ref{eq:averagedlin}) consider the change of coordinates $\tilde{\theta}=\theta-\varphi_0$,
which results in the closed-loop error dynamics
\begin{equation}\label{eq:errvoltclosed}
\begin{gathered}
\frac{d}{dt}\begin{bmatrix}
\tilde{z}_a \\
\tilde{\theta}
\end{bmatrix}=\epsilon\begin{bmatrix}
-E_{md}K_P & E_{md} \\
-K_I & 0
\end{bmatrix}\begin{bmatrix}
\tilde{z}_a \\
\tilde{\theta}
\end{bmatrix}
\end{gathered}
\end{equation}
since $\epsilon$, $E_{md}$ are positive, the matrix in (\ref{eq:errvoltclosed}) is Hurwitz for all $K_P>0$, $K_I>0$, and system (\ref{eq:errvoltclosed}) result asymptotically stable despite the unknown disturbance $\varphi_0$.\\
The problem with implementing the regulator (\ref{eq:errvoltclosed}) is that the resulting control signal is the average value of the actual control input $\eta$, thus some procedure is required to synthesize a real-world control signal whose mean value satisfies the above conditions. In the SAF specific case this problem can be solved, consider the derivative of signal $\eta_a$
\begin{equation}
\frac{d}{dt}\eta_a=\frac{d}{dt}\frac{1}{T}\int_{t-T}^t \eta(\tau)d\tau
\end{equation}
it can be rewritten on the left side as the difference over one period, while the right side is replaced with the derivative of $\eta_a$ expressed in (\ref{eq:PIvolt});
\begin{equation}
\frac{1}{T}[\eta(t)-\eta(t-T)]=-K_P\dot{\tilde{z}}_a+\dot{\tilde{\theta}}=-K_P\dot{\tilde{z}}_a-\epsilon K_I \tilde{z}_a
\end{equation}
solving for $\eta(t)$ yields
\begin{equation}\label{eq:etaimpl}
\eta(t)=-TK_P\dot{\tilde{z}}_a(t)-\epsilon T K_I\tilde{z}_a(t)+\eta(t-T)
\end{equation}
using (\ref{eq:dotza}), the derivative of the averaged square voltage error is actually measurable, thus the above formula is implementable. However, while the interconnection between voltage subsystem and controller is stable in sense of the averaged value, a further step is required. In the incremental implementation (\ref{eq:etaimpl}) there is no more an integral action, the control input history is kept in memory for one period, still the controller provides stability for the averaged voltage error $\tilde{z}_a$. Consider now that for the phasor variables system, a stable steady-state guarantees that all the variables have a constant average value, while being allowed to oscillate freely. This property is desired for what concern the capacitor voltage and it is the main motivation for applying the averaging procedure, however implementation according to (\ref{eq:etaimpl}) can introduce undesired periodic oscillation in the control input $\eta$, moreover oscillation will persist being remembered through the time delay term. In summary, while $\eta_a$ will approach the constant power loss value $\varphi_0$, the actual input $\eta$ might be any periodic signal with average value equal to $\varphi_0$. Recalling that $\eta$ modifies the real power reference value $x_d^*$, any oscillation will result in a non-zero error for the power tracking controller.\\
In order to avoid this situation the following term can be added to (\ref{eq:etaimpl})
\begin{equation}\label{eq:addterm}
d_{\eta}(t)=\eta(t-T)-\eta_a(t-T/2)
\end{equation}
the reason of this modification is to cancel the oscillations stored in memory, by correcting the stored signal towards its own mean value $\eta_a(t-T/2)$. It is important to remark that the averaged value is not the actual mean value of its corresponding signal, the mean value $s_m$ of a signal $s(t)$ is defined as
\begin{equation}
s_m=\frac{1}{T}\int_{t-T/2}^{t+T/2}s(\tau)d\tau
\end{equation}
the above equation is identical to the zero-order phasor definition, except for a time shift of $T/2$.
For this reason the mean value of the stored signal $\eta(t-T)$ has been expressed as its time shifted average value,
note that the mean value of this stored signal can be computed because also its ``future'' values are available.
The additive term $d_{\eta}$ is a zero mean value signal, because it is obtained removing its DC-value to a periodic signal. Since the control input $\eta$ enters the averaged system (\ref{eq:errvoltclosed}) after being averaged itself, any modification having zero mean value will not affect the behavior of the averaged system dynamics. Hence the final implementation of control input together with (\ref{eq:addterm}) is
\begin{equation}\label{eq:etaok}
\eta(t)=-TK_P\dot{\tilde{z}}_a-\epsilon T K_I \tilde{z}_a+\eta_a(t-T/2)
\end{equation}
this controller will not introduce undesired oscillation because it depends solely on averaged signals, whose simplified dynamics (\ref{eq:errvoltclosed}) cannot give oscillations.
\subsection{Power tracking controller}\label{sec:IMC}
The voltage controller output reported in (\ref{eq:etaok}) can be replaced into the filter error power dynamics in (\ref{eq:errdyn}), recalling also equation (\ref{eq:PIvolt}), it turns out
\begin{equation}\label{eq:errdyn_dist}
\begin{aligned}
\dot{\tilde{x}}&=M(R,L)\tilde{x}-\frac{1}{L}\bar{u}+d(t)+f(\epsilon, \tilde{z}_a,\tilde{\theta}, \dot{\tilde{z}}_a, \dot{\tilde{\theta}})\\
\end{aligned}
\end{equation}
where
\begin{equation}
d(t)=d_0+M(R,L)x^*-\dot{x}^*+M(R,L)\varphi_0
\end{equation}
is a T-periodic term composed by the sum of a constant term and sinusoids having known frequency, while
\begin{equation}
\begin{aligned}
f(\tilde{z}_a,\tilde{\theta}, \dot{\tilde{z}}_a, \dot{\tilde{\theta}},\epsilon)=TK_p\ddot{\tilde{z}}_a+\epsilon K_I \dot{\tilde{z}}_a+K_p\dot{\tilde{z}}_a(t-T/2)-\dot{\tilde{\theta}}(t-T/2)\\
+M(R,L)[-T K_p \dot{\tilde{z}}_a-\epsilon K_I\tilde{z}_a-K_p\tilde{z}_a(t-T/2)+\tilde{\theta}(t-T/2)].
\end{aligned}
\end{equation}
The problem of forcing $\tilde{x}$ in (\ref{eq:errdyn_dist}) clearly requires the ability of the control law to compensate for the signal $d(t)$, perfect tracking cannot be achieved by a feedforward action since SAF parameters and $d(t)$ are not fully known. To comply with uncertainties and provide robustness we propose an internal model-based controller. Each component of the vector $d(t)$ can be seen as the output of the following linear system
\begin{equation}\label{eq:esosist}
\begin{aligned}
\dot{w}_i(t)&=\Omega w_i(t), \ \ w_i \in \mathbb{R}^{2N+1}  \\
d_{im}(t)&=\Gamma_i w_i(t), \ \ i=d,q
\end{aligned}
\end{equation}
where $\Gamma_i \in \mathbb{R}^{(1 \times 2N+1)}$ are suitably defined vectors and matrix $\Omega \in \mathbb{R}^{(2N+1)\times(2N+1)}$ is defined as $\Omega=blkdiag(\Omega_j)$ with $\Omega_0=0$ and
\begin{equation}
\Omega_j=\begin{bmatrix}
0 & j\omega_m \\
-j\omega_m & 0
\end{bmatrix}, \ \ j=1,\dots,N
\end{equation}
with the pairs ($\Gamma_i$,$\Omega$) observable. Defining $\Phi=blkdiag(\Omega,\Omega)$ and $\Gamma=blkdiag(\Gamma_d,\Gamma_q)$, the following internal model-based controller can be designed
\begin{equation}\label{eq:imc}
\begin{aligned}
\dot{\xi}=\Phi \xi+Q\tilde{x} \\
\bar{u}=\Gamma \xi +K\tilde{x}
\end{aligned}
\end{equation}
where matrices $Q$ and $K$ need to be properly assigned. Once chosen $\bar{u}$ as in (\ref{eq:imc}) and defined the internal model  error variables as $\tilde{\xi}=\xi-Lw$, where $w:=[w_d^T,w_q^T]^T$, the power subsystem closed-loop error dynamics can be rewritten as
\begin{equation}\label{eq:fastsys}
\begin{aligned}
\dot{\tilde{x}}&=(M(R,L)-\frac{1}{L}K)\tilde{x}-\frac{1}{L}\Gamma \tilde{\xi}+f(\tilde{z}_a,\tilde{\theta}, \dot{\tilde{z}}_a, \dot{\tilde{\theta}},\epsilon)\\
\dot{\tilde{\xi}}&=\Phi \tilde{\xi}+Q\tilde{x}. \\
\end{aligned}
\end{equation}
According to the general two time-scale averaging theory, the power tracking problem can be studied focusing on the \emph{boundary layer system}, obtained by putting $\epsilon=0$ into the overall error dynamics, hence by (\ref{eq:zanou}), (\ref{eq:PIvolt}) and $\dot{\tilde{z}}_a=0$, $\dot{\tilde{\theta}}=0$, thus system (\ref{eq:fastsys}) becomes
\begin{equation}\label{eq:boundlayer}
\begin{aligned}
\dot{\tilde{x}}&=(M(R,L)-\frac{1}{L}K)\tilde{x}-\frac{1}{L}\Gamma \tilde{\xi}+f(\tilde{z}_a,\tilde{\theta}, 0, 0, 0)\\
\dot{\tilde{\xi}}&=\Phi \tilde{\xi}+Q\tilde{x}.
\end{aligned}
\end{equation}
Now matrices $K$, $Q$  need to be selected such that asymptotic stability is provided for the boundary layer system. Define two arbitrary Hurwitz matrices $F_d$, $F_q$  $\in \mathbb{R}(2N+1)\times(2N+1)$, and two arbitrary vectors $G_d$, $G_q$ such that the pairs ($F_d$, $G_d$), ($F_q$, $G_q$) are controllable, taking the controller matrices as
\begin{equation}\label{eq:matrimc}
\begin{gathered}
K=k\begin{bmatrix}
k_d & 0 \\
0 & k_q
\end{bmatrix}, \ Q=
\begin{bmatrix}
E_d^{-1} & 0 \\
0 & E_q^{-1}
\end{bmatrix}
\begin{bmatrix}
G_d & 0 \\
0 & G_q
\end{bmatrix}K
\end{gathered}
\end{equation}
with $k_d$, $k_q$ two arbitrary positive scalars, $k$ a positive design parameter, and $E_d$, $E_q$ defined as non-singular solutions of the following Sylvester equations:
\begin{equation}\label{eq:eqsylv}
\begin{aligned}
F_dE_d-E_d\Omega_d&=-G_d\Gamma_d \\
F_qE_q-E_q\Omega_q&=-G_q\Gamma_q \\
\end{aligned}
\end{equation}
asymptotic stability of the boundary layer system can be stated. In order to prove this claim let us define the vector
\begin{equation}\label{eq:Rxi}
R_{\xi}=\left[ -\frac{R}{\Gamma_{d1}} \ \ 0_{2N} \ \ -\frac{\omega_mL}{\Gamma_{q1}} \ \ 0_{2N}  \right]^T
\end{equation}
where $\Gamma_{d1}$, $\Gamma_{q1}$ denote the first element of vectors $\Gamma_d$, $\Gamma_q$ respectively and $0_{2N}$ is a zero raw vector having dimension $2N$. Consider now the change of variables
\begin{equation}
\tilde{\chi}=E\tilde{\xi}-ER_{\xi}(\tilde{\theta}(t-T/2)-K_p\tilde{z}_a(t-T/2))+LG\tilde{x}
\end{equation}
where $E=blkdiag(E_d,E_q)$, $G=blkdiag(G_d,G_q)$, in this coordinates system (\ref{eq:boundlayer}) results
\begin{equation}
\begin{aligned}
\dot{\tilde{x}}&=(M(R,L)-\frac{1}{L}K +\Gamma L^{-1}G)\tilde{x}-\frac{1}{L}\Gamma E^{-1} \tilde{\chi}\\
\dot{\tilde{\chi}}&=F \tilde{\chi}-L(FG-GM(R,L))\tilde{x} \\
\end{aligned}
\end{equation}
where $F=blkdiag(F_d,F_q)$. Using standard linear system tools it can be verified that a value $\bar{k}$ exists, such that $\forall \ k\geq \bar{k}$ the state matrix of the system in the new coordinates is Hurwitz, hence asymptotic stability of the boundary layer system can be stated.
\subsection{Overall system stability}
Asymptotic stability has been stated for the boundary layer system and a linearized version of the averaged reduced voltage dynamics. Exploiting  the main results of the two time-scale averaged systems theory, it can be proved that the two separately designed power and DC-bus voltage controllers, are able to provide practical stability for the complete system (\ref{eq:errdyn}),  that is it's possible to claim that there exists a value $\epsilon^*$, such that $\forall \epsilon<\epsilon^*$, $k\geq \bar{k}$, $l \leq l^*$ the set
\begin{equation*}
\{(\tilde{x},\tilde{\xi}):\tilde{x}=0, \ \tilde{\xi}=0\}\times \{(\tilde{z},\tilde{\theta}): |\tilde{z}|\leq l^*, \tilde{\theta}=0 \}
\end{equation*}
is \emph{practically stable} \citep{Khalil96} for the closed-loop trajectories of the complete error system. The fact that the proposed regulator is able to achieve the control objectives in a practical way means that the power vector $x$ can be steered arbitrary close to the reference value $x_{\varphi_0}^*$ while the averaged value $\eta_a$ tends arbitrary close to the power loss term $\varphi_0$. It's further to notice that the asymptotic tracking error can be arbitrary reduced by taking a smaller value for $\epsilon$, that is by increasing the capacitor value $C$.
\section{Simulation results} \label{sec:simulazioni}
\begin{figure}[H]	
\centering
\psfragscanon
\subfigure[Real power component tracking error.]
{\includegraphics[width=5cm, height=4cm]{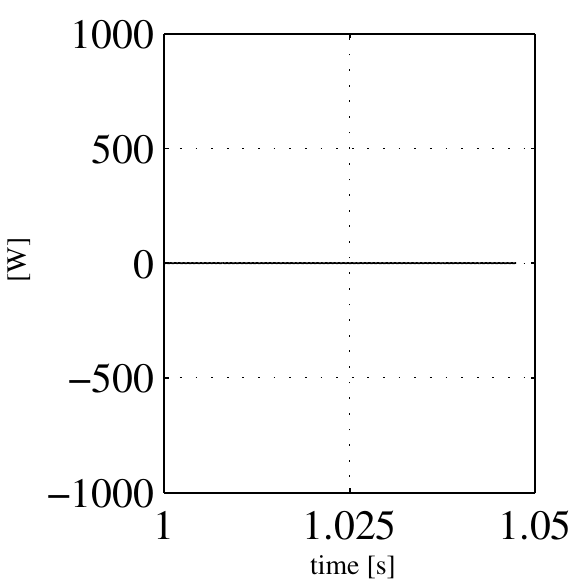}} 
\subfigure[Virtual power component tracking error.]
{\includegraphics[width=5cm, height=4cm]{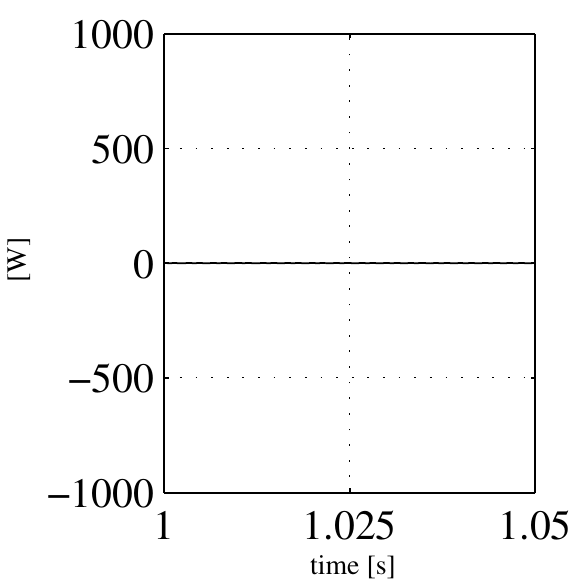}} 
\caption{Error variables $\tilde{x}_d$, $\tilde{x}_q$: two harmonics load scenario.}\label{fig:xtildeI}
\end{figure}

\begin{figure}[H]	
\centering
\psfragscanon
\includegraphics[width=5cm, height=4cm]{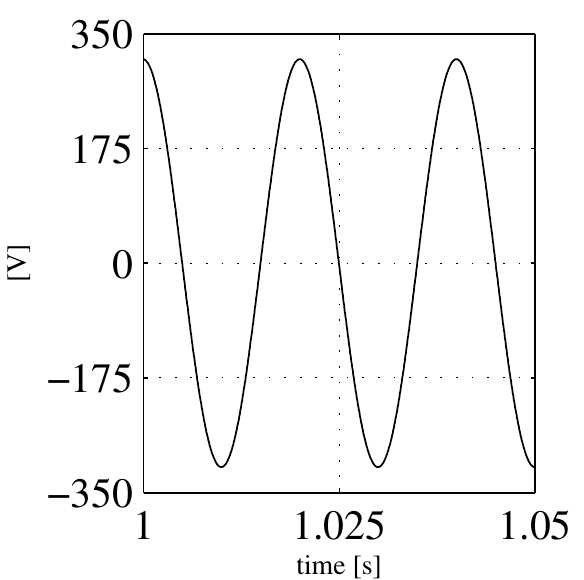}
\includegraphics[width=5cm, height=4cm]{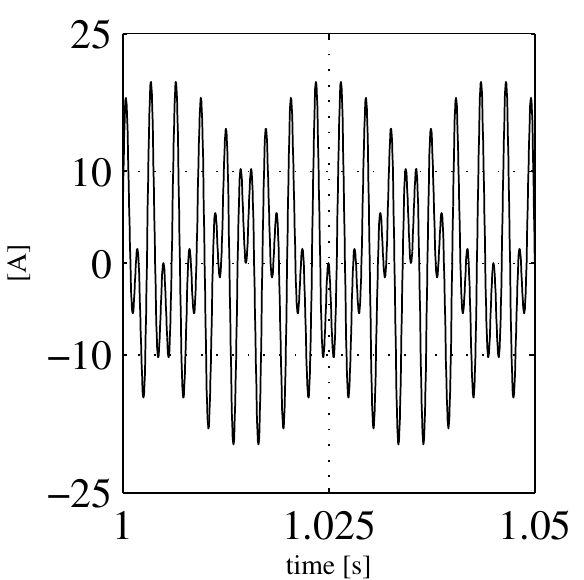} \\
\includegraphics[width=5cm, height=4cm]{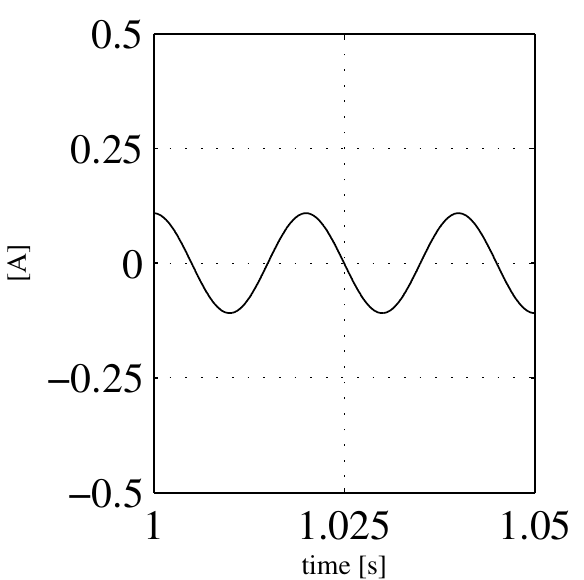}
\includegraphics[width=5cm, height=4cm]{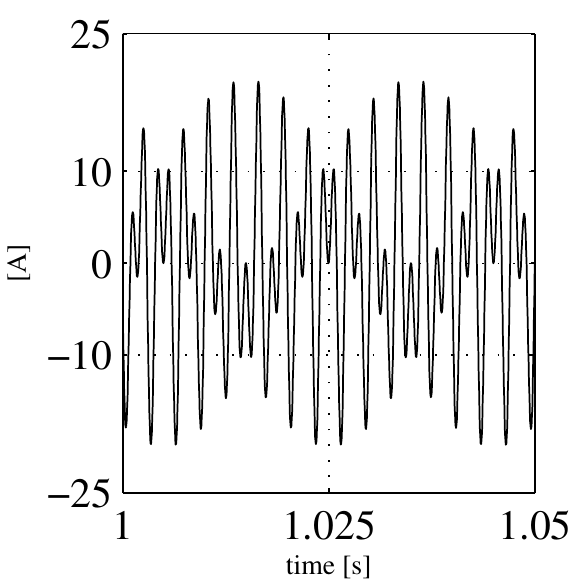}
\caption{Current and line voltage waveforms on phase $a$: two harmonics load scenario.}\label{fig:imfl_Ideal}
\end{figure}

\begin{figure}[H]	
\centering
\psfragscanon
\subfigure[Main current magnitude spectrum.]
{\includegraphics[width=5cm, height=4cm]{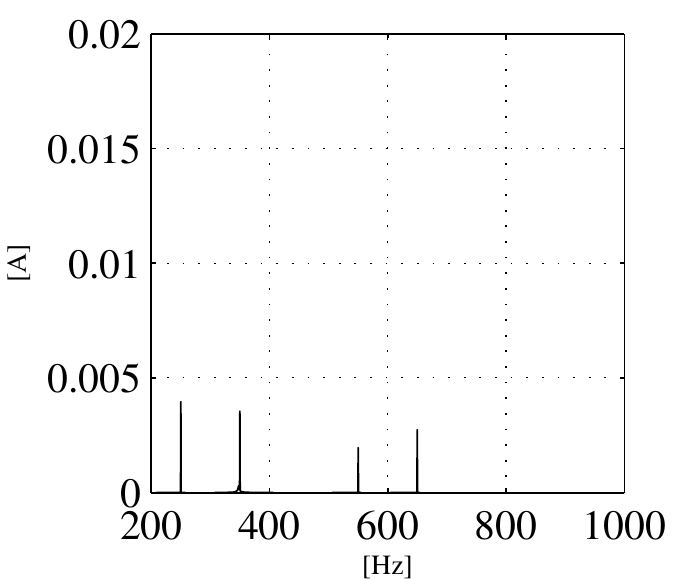}}
\subfigure[Load current magnitude spectrum.]
{\includegraphics[width=5cm, height=4cm]{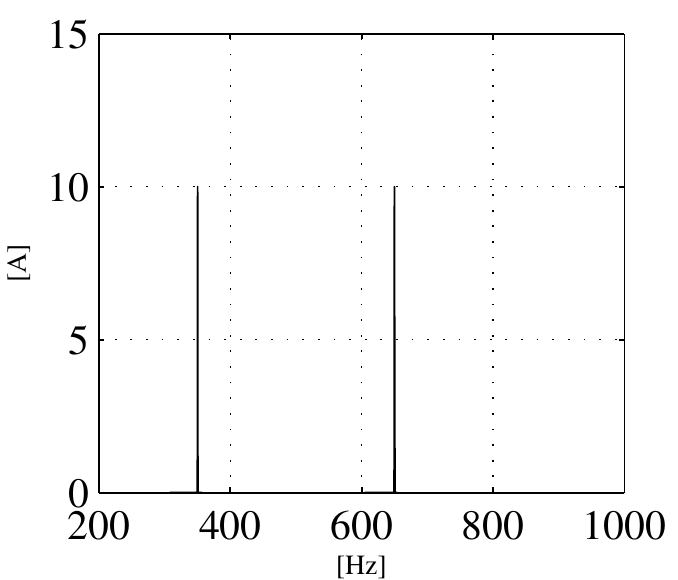}}
\caption{FFT of the a-phase main current and of the corresponding load current: two harmonics load scenario.}\label{fig:spettro_Ideal}
\end{figure}

\begin{figure}[H]	
\centering
\psfragscanon
\subfigure[Square capacitor voltage error and computed average value (bold).]
{\includegraphics[width=5cm, height=4.3cm]{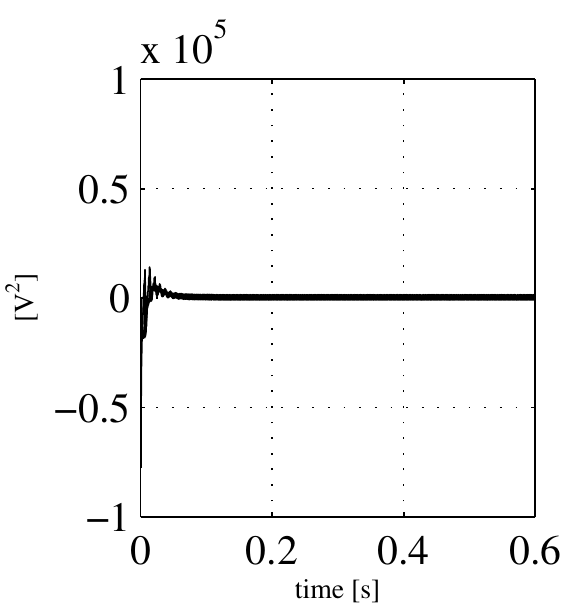}}
\subfigure[Actual capacitor voltage value.]
{\includegraphics[width=5cm,height=4cm]{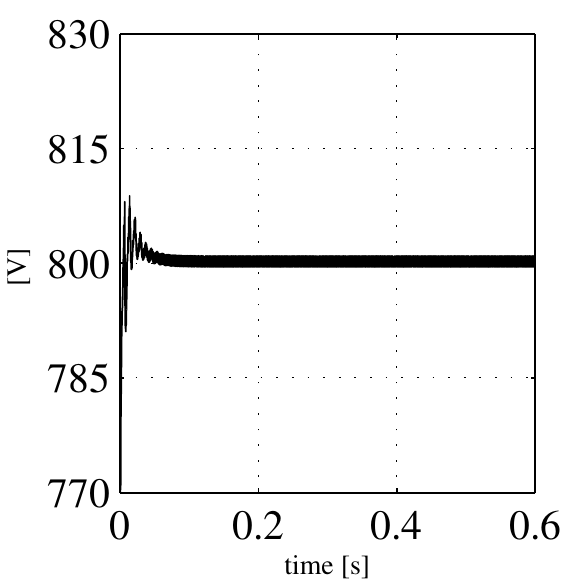}}
\caption{Voltage controller performance: two harmonics load scenario.}\label{fig:voltstabIdeal}
\end{figure}

\begin{figure}[H]	
\centering
\psfragscanon
\subfigure[Real power component tracking error.]
{\includegraphics[width=5cm,height=4cm]{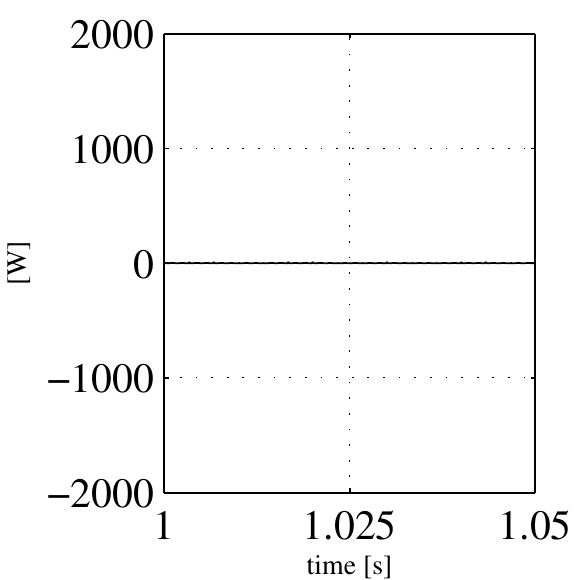}} 
\subfigure[Virtual power component tracking error.]
{\includegraphics[width=5cm,height=4cm]{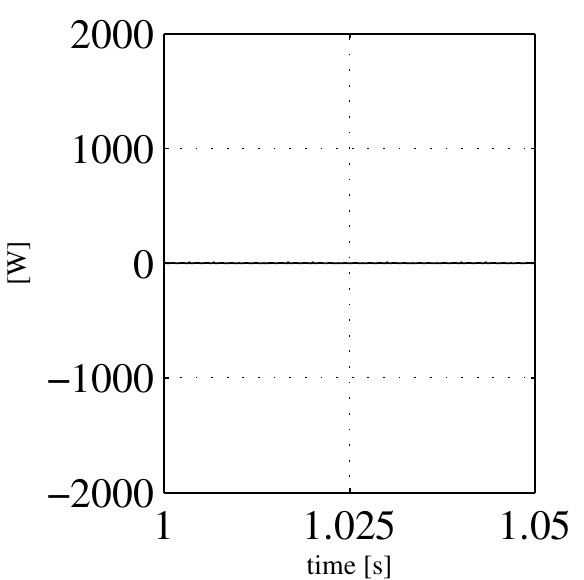}} 
\caption{Error variables $\tilde{x}_d$, $\tilde{x}_q$: diode bridge load scenario.}\label{fig:xtildeR}
\end{figure}

\begin{figure}[H]	
\centering
\psfragscanon
\includegraphics[width=5cm, height=4cm]{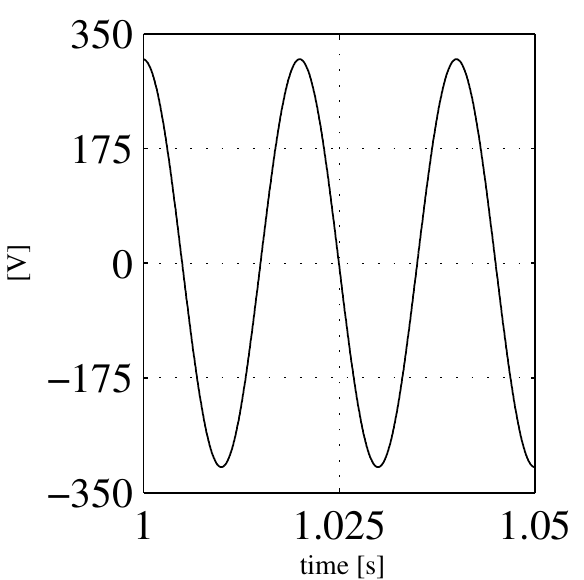}
\includegraphics[width=5cm, height=4cm]{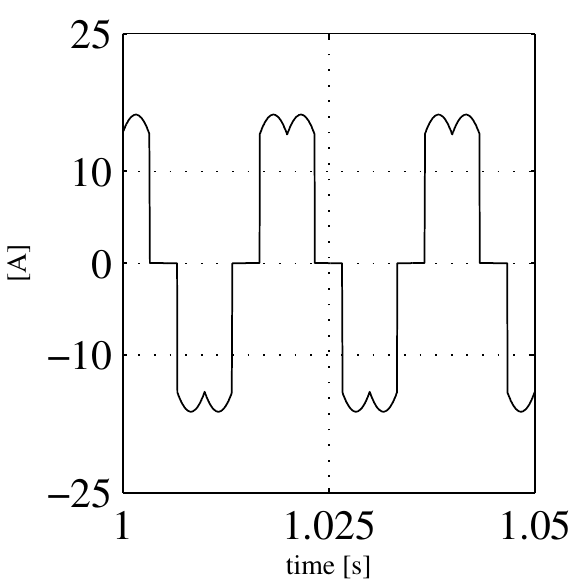} \\
\includegraphics[width=5cm, height=4cm]{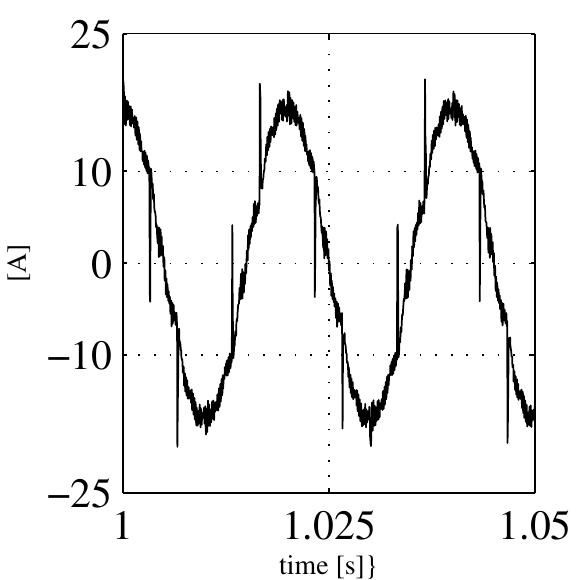}
\includegraphics[width=5cm, height=4cm]{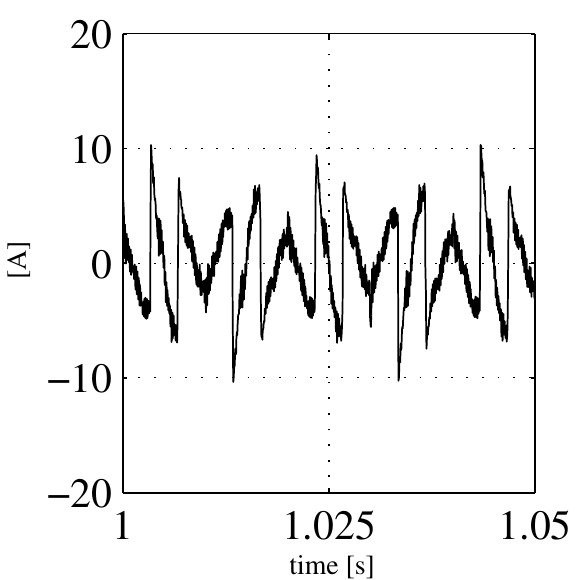}
\caption{Current and line voltage waveforms on phase $a$: diode bridge load scenario.}\label{fig:imfl_Real}
\end{figure}

\begin{figure}[H]	
\centering
\psfragscanon
\subfigure[Main current magnitude spectrum.]
{\includegraphics[width=5cm, height=4cm]{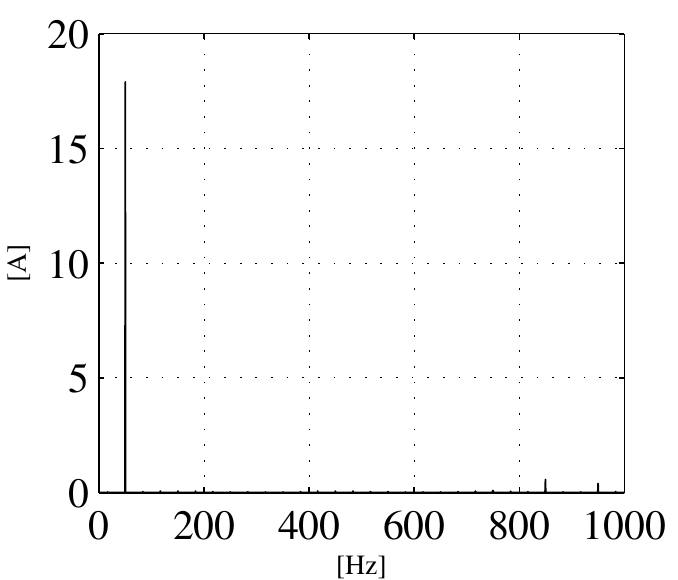}}
\subfigure[Load current magnitude spectrum.]
{\includegraphics[width=5cm, height=4cm]{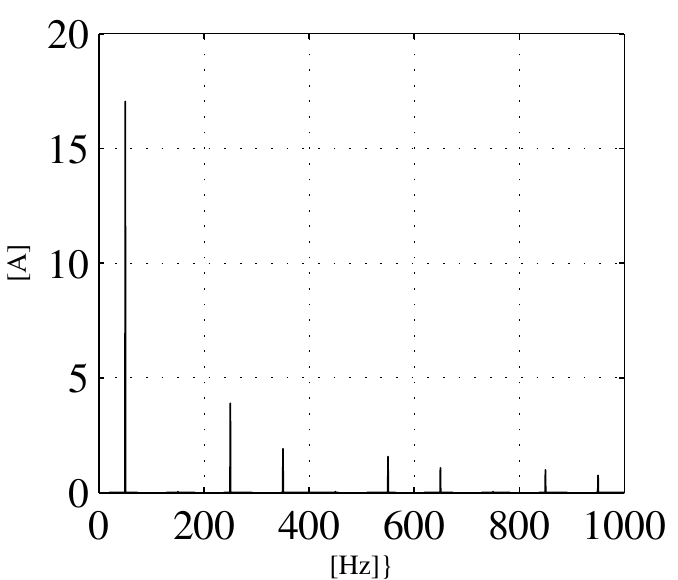}}
\caption{FFT of the a-phase main current and of the corresponding load current: diode bridge load scenario.}\label{fig:spettro_Real}
\end{figure}

\begin{figure}[H]	
\centering
\psfragscanon
\subfigure[Square capacitor voltage error and computed average value (bold).]
{\includegraphics[width=5cm,height=4.3cm]{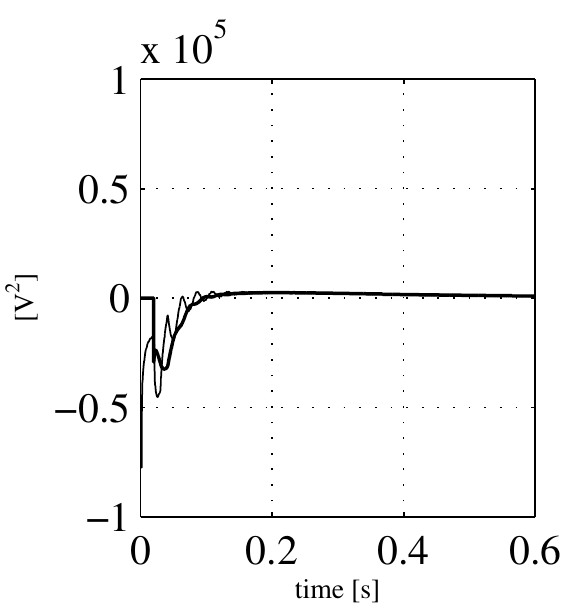}}
\subfigure[Actual capacitor voltage value.]
{\includegraphics[width=5cm,height =4cm]{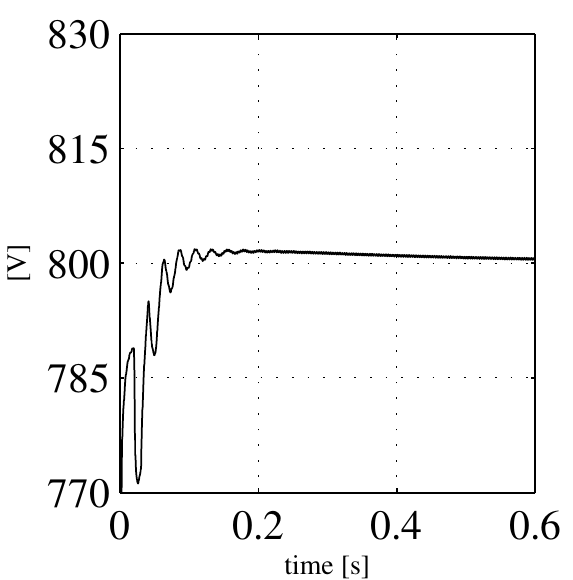}}
\caption{Voltage controller performance: diode bridge load scenario.}\label{fig:voltstabReal}
\end{figure}

Simulation tests have been performed in order to validate the proposed control solution. Two different scenarios have been adopted; first model (\ref{eq:modabc_comp}) has been implemented in MATLAB/Simulink and a load scenario with two harmonics at $7\omega_m$ and $13\omega_m$ has been chosen. Then, in order to validate the controller performance in a situation closer to a real setup, the proposed continuous-time regulator has been discretized adopting a sampling frequency $f_s=7$ $KHz$, then the SAF converter components have been modeled by using Simulink/SimPowerSystems toolbox, and  a suitable PWM technique with a carrier frequency equal to $f_s$ has been implemented. Finally a three phase diode bridge has been selected as nonlinear load scenario.\\
The following system parameters has been set, according to the procedure illustrated in Section \ref{sec:dimensionamento}; $C=4400$$\mu F$, $L=3.3$$mH$, $R=0.12 \Omega$, while the DC-link voltage limits have been set to $v_m=700V$, $v_M=900V$. Ideal three-phase mains voltages with amplitude $V_m=310$$V$ and frequency $f_m=50$$Hz$ have been modeled.\\
The internal-model based controller has been tuned to the load disturbances, according to the procedure described in \ref{sec:IMC}, for what concern the simulations in time continuous domain. As regards the diode rectifier load scenario, the most relevant power disturbances, that is the $6^{th}$ and the $12^{th}$ load current harmonics expressed in the \emph{synchronous} $d-q$ reference frame (corresponding respectively to the $5^{th}$ and the $7^{th}$, and to the $11^{th}$ and the $13^{th}$ in the fixed reference frame), have been considered, then the IMC controller has been discretized according to the procedure reported in \citep{Ronchi2003}, thus the following matrices have been selected; $\Omega=blkdiag(\Omega_0,\Omega_6,  \Omega_{12})$, $\Gamma_d=\Gamma_q=(1,1,0,1,0)^T$, $K=diag(200,200)$ and $Q=10^3diag(Q_d,Q_q)$, where $Q_d=Q_q=(40.6,80.7,7.15,78.7,17.6)^T$. For what concerns the voltage stabilizer described in \ref{sec:AveragingV}, the following parameters have been selected $K_P=0.3$, $K_I=3.7$.\\
Consider now the performance obtained on the first simulation scenario, with ideal SAF model and the $7^{th}$, $13^{th}$  disturbance harmonics; in Fig. \ref{fig:xtildeI} the tracking error on both real and imaginary power variables is reported, as expected, asymptotic perfect tracking is achieved and the vector $\tilde{x}$ is steered to the origin. This ideal behavior is confirmed by Fig. \ref{fig:imfl_Ideal}, \ref{fig:spettro_Ideal}; the two harmonics currents are totally canceled out by the filter currents, while a small current component oscillating at the first-order harmonic frequency and aligned to the corresponding voltage, arises on the line side due to the voltage controller action. In table \ref{tab:perfId} the harmonics compensation performance are summarized.\\
\begin{table}[H]
\centering
\begin{tabular}{|c|c|c|c|} \hline
Harmonic frequency [Hz] & $i_{ma}$ [A]& $i_{la}$ [A] & Compensation percentage \\  \hline
350 & 0.0039 & 10  & 99.96\%     \\ \hline
650 & 0.0038 & 10 &  99.96\%   \\ \hline
\end{tabular}
\caption{Compensation performance for the two harmonics disturbance scenario.}\label{tab:perfId}
\end{table}
For what concerns the voltage controller, in order to validate the stability properties, a value quite far from the mean voltage reference value $(v_m^2+v_M^2)/2=800$ $V$ has been chosen as initial condition for the capacitor voltage. As showed in Fig. \ref{fig:voltstabIdeal}, even though the average value is initialized at zero and needs one period before representing the actual voltage, the voltage controller reacts immediately, thanks to its dependance on the averaged derivative $\dot{\tilde{z}}_a$. Hence the voltage averaged error is successfully steered to zero, and
the capacitor voltage is brought back to the middle of the safe interval, without exceeding the upper and lower bounds. The initial nonlinear behavior of the voltage error trajectories is originated by the neglected nonlinearities and also by the coupling term $\tilde{D}(\tilde{x})$, although it has been neglected due to two time-scale behavior hypothesis, it's excited by the internal model controller transient when harmonics compensation starts.\\
As regards the second simulation scenario, carried out in discrete time domain and with a more detailed filter physical model, the power tracking performance are reported in Fig. \ref{fig:xtildeR}, \ref{fig:imfl_Real}, \ref{fig:spettro_Real}, in this case the power error variables $\tilde{x}_d$, $\tilde{x}_q$ are not exactly zero, due to the fact that AC/DC rectifier high order harmonics are not compensated by the internal model, furthermore the discretization effects have to be taken into account. However the load currents harmonics for which the controller has been tuned are strongly reduced at the line side as the currents magnitude spectrum reported in Fig. \ref{fig:spettro_Real} shows.
Analyzing the currents waveform in the time domain (Fig. \ref{fig:imfl_Real}), it can be verified that the mains currents are almost sinusoidal and perfectly aligned with the corresponding phase voltages, hence also the load imaginary power is almost totally compensated. The ripple introduced by the pulse with modulation can be noted on the filter current, it affects also the mains currents, however thanks to a correct inductance sizing, the high frequency distortion is properly bounded.
Quantitative performance of the power-tracking controller obtained with this scenario are summarized in Tab. \ref{tab:perfRe}.\\
\begin{table}[t]
\centering
\begin{tabular}{|c|c|c|c|} \hline
Harmonic frequency [Hz] & $i_{ma}$ [A] & $i_{la}$ [A] & Compensation percentage \\  \hline
250 & 0.03 & 3.88 &  99.2\%  \\ \hline
350 & 0.04 & 1.91 &  97.9\% \\ \hline
550 & 0.03 & 1.57 &  98.1\%  \\ \hline
650 & 0.02 & 1.08 &  98.1\%  \\ \hline
\end{tabular}
\caption{Compensation performance for the diode bridge load scenario.}\label{tab:perfRe}
\end{table}
The current component corresponding to the line frequency oscillation is slightly larger at the line side than at the load side, due to the additional active power drained to compensate for the filter losses.\\
As regards the averaging voltage controller, a discrete time version has been implemented, while the same initial conditions of the first scenario have been reproduced. In Fig. \ref{fig:voltstabReal} the squared voltage error, it's averaged value and the actual capacitor voltage are reported, also in this case the objective relative to the voltage dynamics behavior is accomplished, similar considerations to those made for the previous scenario can be made.
\section{Conclusions} \label{sec:conclusioni}
In this chapter a nonlinear robust control solution for a shunt active filter has been proposed, the focus has been firstly put on the hardware components design issue, providing a suitable algorithm, based on the structural system properties,  which gives guarantees on the feasibility of the control problem and allows to obtain a crucial time-scale separation between the power and voltage dynamics. Then exploiting nonlinear systems analysis well established tools, such as averaging and singular perturbation theory, an averaging capacitor voltage controller and a power tracking controller based on the internal model principle, have been presented. The former exploits the insight that, regulating the averaged voltage value, makes it possible to ignore the necessary oscillations for a proper filter operation, and improves the voltage dynamics behavior. The second is chosen in order to ensure asymptotic tracking of undesired load current components, providing also robustness with respect to disturbances and model uncertainties.\\
Saturation issues have not been explicitly addressed in this work, owing to space limitation, however it is of utmost importance to deal with these phenomena for an actual industrial implementation with stability and performance guarantees. Some solutions, for the SAF specific case, have been proposed (see \cite{Cavini2004a}, \cite{Cavini2004b}), however this is still an open research topic. Future effort will thus be devoted to improve the filter performance under control input saturation, analyzing the problem in the context of modern anti-windup approaches, hence providing a rigorous characterization of the system under saturation constraints. Moreover discretization issues relative to the nonlinear controller here discussed will be further analyzed, in order to improve the discrete-time controller performance with respect to that obtained applying standard discretization techniques.

\end{document}